\documentclass[%
 reprint,
 amsmath,amssymb,
 aps,
]{revtex4-2}
\usepackage{xcolor}

\usepackage{graphicx}
\usepackage{dcolumn}
\usepackage{bm}

\usepackage{braket}
\usepackage{subcaption}
\usepackage{hyperref}
\begin{document}

\preprint{APS/123-QED}

\title{Structured-Light Magnetometry in a Coherently Controlled Atomic Medium}

\author{Parkhi Bhardwaj}
\email{parkhi.21phz0013@iitrpr.ac.in} 
\author{Shubhrangshu Dasgupta}%
 \affiliation{%
 Department of Physics, Indian Institute of Technology Ropar, Rupnagar,  Punjab 140001, India
}%




\date{\today}

\begin{abstract}
 A structured-light-based approach to magneto-optical rotation (MOR) magnetometry is proposed, in which magnetically induced polarization rotation is converted into a directly observable spatial displacement. A radially polarized Laguerre--Gaussian beam propagating through cold $^{87}\mathrm{Rb}$ atoms in the presence of a longitudinal magnetic field experiences circular birefringence, which induces a relative phase shift between the $\sigma_+$ and $\sigma_-$ polarization components. This phase shift manifests itself as the angular rotation of a petal-shaped interference pattern, enabling the MOR angle to be determined directly from the spatial intensity distribution. Unlike conventional MOR detection techniques based on balanced polarimetry or Stokes-parameter reconstruction, the proposed method requires only a fixed-axis polarization projection and avoids computationally intensive polarization analysis. By translating polarization rotation into a topology-dependent spatial observable, the scheme enables direct and spatially resolved magnetic-field measurements. Furthermore, the use of electromagnetically induced transparency (EIT) allows low-loss operation while preserving high sensitivity. These characteristics make the proposed technique a promising platform for optical magnetometry and structured-light-based quantum sensing.
\end{abstract}

\maketitle


\section{\label{sec:level1} Introduction}
Magnetometry, the precise measurement of magnetic fields \cite{10.1116/5.0025186}, is essential in a wide range of applications, including geophysics \cite{Bennett2021-im}, biomedical imaging \cite{Aslam2023}, materials characterization \cite{10.3389/fphy.2023.1212368}, quantum information processing \cite{RevModPhys.92.015004, AN2022103752}, navigation \cite{Bennett2021-im}, nondestructive testing \cite{10963667}, and magnetoencephalography (MEG) \cite{10.1063/1.5001730}. The ability to detect weak magnetic fields with high spatial and temporal resolution is crucial for both fundamental studies and technological applications.

Conventional magnetometers, including fluxgate sensors \cite{s21041500}, Hall probes \cite{10.1063/1.1717911}, and superconducting quantum interference devices (SQUIDs) \cite{133898}, are widely employed for magnetic-field sensing. However, their performance is often limited by requirements such as cryogenic operation, frequent calibration, or restricted bandwidth \cite{Wei2021-cl,10.1063/1.2354545}. In contrast, atomic magnetometers exploit magnetic-field-dependent atomic dynamics to convert Zeeman shifts into measurable optical signals, including changes in absorption, fluorescence, and polarization rotation \cite{10.1063/1.3491215,10.3389/fphy.2023.1212368}.

Among these techniques, magneto-optical-rotation (MOR)-based magnetometers have attracted considerable attention because of their all-optical and noninvasive nature \cite{Fabricant_2023}. In such systems, a longitudinal magnetic field induces polarization rotation of the probe field, from which the magnetic-field strength can be determined \cite{10.3389/fphy.2022.946515,CHEN2023100152}. Their sensitivity is primarily limited by photon shot noise and atomic decoherence \cite{Fabricant_2023,PhysRevLett.113.013001,PhysRevA.82.033807,PhysRevLett.97.230801,PhysRevA.93.063826}.

Recently, structured light has emerged as a promising tool for extending MOR-based magnetometry beyond conventional Gaussian beams. Optical fields carrying spatially varying polarization distributions or orbital angular momentum (OAM), often combined with electromagnetically induced transparency (EIT) or nonlinear optical processes, can enhance light--matter interactions and improve magnetic-field sensitivity \cite{Sun:23,https://doi.org/10.1002/lpor.202400465,Chatterjee:25,10.1063/1.4923446,Qiu:21,Fang:25,TAMBAG2023129649,Ghaderi_Goran_Abad2021-jc,Daloi:22,PhysRevA.105.063714,Bhardwaj2025}. Nevertheless, most existing schemes still rely on conventional polarization analysis to extract the MOR angle.

MOR is commonly detected using balanced polarimetry \cite{Patterson2015,71503} or polarization-resolved imaging techniques \cite{TIWARI2025109427}. Such approaches generally require either careful polarization-element alignment or Stokes--Mueller analysis involving matrix reconstruction \cite{Chang:14}. Structured-light-based methods have also enabled magnetic-field imaging, although many rely on absorption measurements or polarization-state reconstruction \cite{PhysRevLett.127.233202,202400465,Cai:26,b47h-pdv9,Chang:14}. In contrast, the present approach encodes the magnetically induced polarization rotation directly into the spatial intensity distribution, allowing the MOR angle to be extracted from the angular position of the intensity pattern without full Stokes-vector reconstruction. Here, EIT-induced circular birefringence is exploited in a low-loss structured-light magnetometer based on radially polarized Laguerre--Gaussian (LG) beams. The common-path configuration provides a simple single-beam readout, while an extended two-beam configuration allows the number of interference petals, and hence the spatial resolution, to be controlled through the OAM topological charges.

Conventional atomic magnetometers based on Faraday rotation generally employ balanced photodetection and electronic demodulation techniques and are capable of achieving femtotesla sensitivities in the spin-exchange relaxation-free (SERF) regime under highly controlled conditions requiring stringent magnetic shielding and ultrahigh atomic densities \cite{8425967}. In contrast, the proposed scheme operates at finite magnetic fields and does not require either extreme shielding conditions or exceptionally high atomic densities. Instead, it converts the magnetic-field-dependent circular birefringence of the atomic medium directly into the spatial rotation of an interference pattern, enabling alignment-free, real-time, and spatially resolved magnetic-field measurements.

The estimated sensitivity of the proposed method lies in the nT$/\sqrt{\mathrm{Hz}}$ range, making it comparable to that of fluxgate magnetometers \cite{s21041500}. Furthermore, the sensitivity can be optimized by varying the control-field strength and the interaction length, since the accumulated Faraday rotation increases with the optical depth of the medium. Compared with fluxgate magnetometers, the present technique provides a completely optical platform capable of simultaneously performing quantitative magnetic-field measurements and direct visualization of magnetic-field distributions. Moreover, unlike NV-center-based magnetometers, the proposed method offers a macroscopic platform that does not rely on solid-state nanostructures \cite{RevModPhys.92.015004}.

The remainder of this paper is organized as follows. In Sec.~II, we present the theoretical model describing the interaction of the electromagnetic (EM) field with the atomic medium in the presence of a static magnetic field. In Sec.~III, we introduce two interference-based detection schemes, for measuring the MOR angle using radially polarized LG fields. Section~IV is devoted to the analysis of the principal noise sources and the optimization of the magnetometer sensitivity. Finally, the main conclusions are summarized in Sec.~V.


\section{\label{sec:level2} Theoretical Model}
 The atomic system considered here is an ensemble of cold $^{87}$Rb atoms in a tripod-type four-level configuration realized using Zeeman sublevels of the D$_2$ line in the presence of a homogeneous static magnetic field applied along the $z$-direction. The three ground states as shown in Fig.\ref{1a} are denoted by $|1\rangle \equiv |5S_{1/2},F=1,m_F=-1\rangle$, $|2\rangle \equiv |5S_{1/2},F=1,m_F=0\rangle$, and $|3\rangle \equiv |5S_{1/2},F=1,m_F=+1\rangle$, while the excited state is $|4\rangle \equiv |5P_{3/2},F'=0,m_F'=0\rangle$ {\cite{2175-ztdj}. The magnetic field lifts the degeneracy of the ground-state sublevels through Zeeman splitting.
A radially polarized probe field couples to the $D_2$ transition, where its $\sigma_+$ and $\sigma_-$ components drive the transitions $\ket{4} \leftrightarrow \ket{1}$ and $\ket{4} \leftrightarrow \ket{3}$ with Rabi frequencies $\Omega_+ = \vec{d}_{41}\cdot \vec{E}_+/\hbar$ and $\Omega_- = \vec{d}_{43}\cdot \vec{E}_-/\hbar$, respectively. A \(\pi\) polarized control field couples the transition $\ket{4} \leftrightarrow \ket{2}$ with Rabi frequency $\Omega_c = \vec{d}_{42}\cdot \vec{E}_c/\hbar$.The spontaneous decay channels $|4\rangle \rightarrow |i\rangle$ are characterized by the respective rates $\gamma_{4i}$ $(i\in 1,2,3)$.

\begin{figure}
\centering

    \includegraphics[width=7cm]{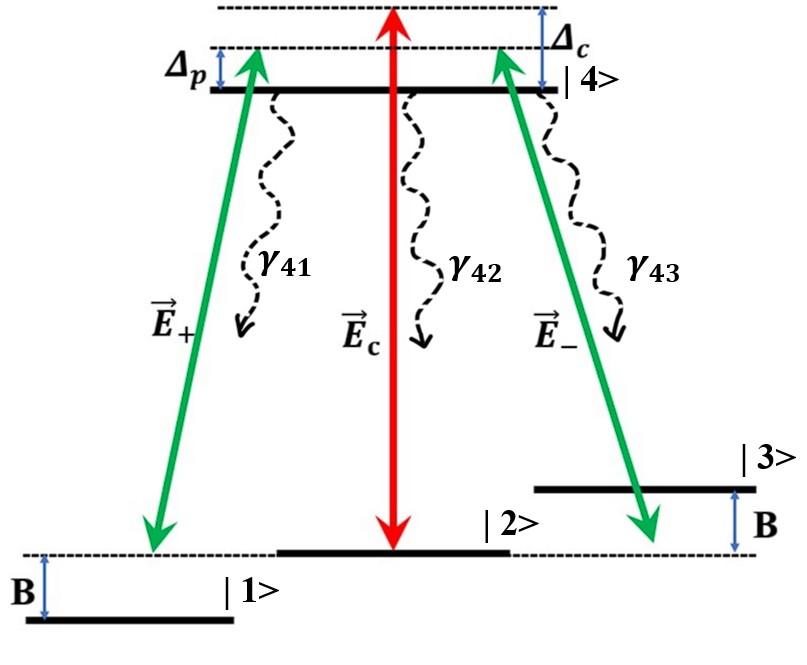}
    \caption{(a) Energy-level configuration of the four-level atomic system interacting with the optical fields. }
\label{1a}
\end{figure}

Within the rotating-wave approximation, the system Hamiltonian is written as  \cite{Scully_Zubairy_1997, RevModPhys.77.633}
\begin{equation}
\begin{aligned}
\frac{\mathcal{H}}{\hbar} &=
- \Big( B |1\rangle\langle 1|
+ (\Delta_p - \Delta_c)|2\rangle\langle 2|
- B |3\rangle\langle 3| + \Delta_p |4\rangle\langle 4|\Big) \\
& \quad +\Big( \Omega_+ \ket{4}\bra{1}
+ \Omega_c \ket{4}\bra{2}
+ \Omega_- \ket{4}\bra{3} +{\rm h.c.}\Big),
\end{aligned}
\end{equation}  
where $\Delta_p$ and $\Delta_c$ are the probe and control detunings, respectively, and $B$ represents the Zeeman shift due to the applied magnetic field.  

The atomic dynamics are governed by the Liouville equation for the density operator $\rho$ \cite{Scully_Zubairy_1997}. The relevant equations of motion are  
\begin{equation}
\begin{aligned}
\dot{\rho}_{11} &= i \left( \Omega_+^* \rho_{14} - \Omega_+ \rho_{41} \right) + \gamma_{41} \rho_{44}, \\
\dot{\rho}_{22} &= i \left( \Omega_c^* \rho_{42} - \Omega_c \rho_{24} \right) + \gamma_{42} \rho_{44}, \\
\dot{\rho}_{44} &= i \big( \Omega_+ \rho_{14} - \Omega_+^* \rho_{41} 
+ \Omega_c \rho_{24} - \Omega_c^* \rho_{42} 
+ \Omega_- \rho_{34} - \Omega_-^* \rho_{43} \big) \\
& \quad - (\gamma_{41} + \gamma_{42} + \gamma_{43})\rho_{44}, \\
\dot{\rho}_{41} &= i \Delta_{14}\rho_{41} + i \left\{ \rho_{31}\Omega_- + \rho_{21}\Omega_c + (\rho_{11}-\rho_{44})\Omega_+ \right\}, \\
\dot{\rho}_{42} &= i \Delta_{24}\rho_{42} + i \left\{ \rho_{32}\Omega_- + \rho_{12}\Omega_+ + (\rho_{22}-\rho_{44})\Omega_c \right\}, \\
\dot{\rho}_{43} &= i \Delta_{34}\rho_{43} + i \left\{ \rho_{13}\Omega_+ + \rho_{23}\Omega_c + (\rho_{33}-\rho_{44})\Omega_- \right\}, \\
\dot{\rho}_{12} &= i \Delta_{12}\rho_{12} + i \left( \rho_{42}\Omega_+^* - \rho_{14}\Omega_c \right), \\
\dot{\rho}_{13} &= i \Delta_{13}\rho_{13} + i \left( \rho_{43}\Omega_+^* - \rho_{14}\Omega_- \right), \\
\dot{\rho}_{23} &= i \Delta_{23}\rho_{23} + i \left( \rho_{42}\Omega_c^* - \rho_{24}\Omega_- \right).
\end{aligned}
\end{equation}  

The detunings are defined as $\Delta_{j4} = \Delta_p \mp B + i\Gamma_{j4}$ ($j\in 1,3$), $\Delta_{24} = \Delta_c + i\Gamma_{24}$, $\Delta_{12} = B - (\Delta_p - \Delta_c) + i\Gamma_{12}$, $\Delta_{13} = 2B + i\Gamma_{13}$, and $\Delta_{23} = B + (\Delta_p - \Delta_c) + i\Gamma_{23}$. Here $\Gamma_{ij}$ are decoherence rates and $\gamma_{ij}$ represent spontaneous decay rates from the level $|i\rangle$ to $|j\rangle$.  

In the steady-state regime, under the weak-probe approximation ($\Omega_\pm \ll \Omega_c$), the equations are solved perturbatively up to first order in the probe field while retaining all orders in the control field \cite{RevModPhys.77.633}. In this limit, the populations are approximated as $\rho_{11} = \rho_{33} = 1/2$ and $\rho_{22} = \rho_{44} = 0$. The probe-induced coherences are then given by $\rho_{41}=\Omega_+\rho_{41}^{(1)},\rho_{43}=\Omega_-\rho_{43}^{(1)}$, where
\begin{equation}
    \rho_{41}^{(1)} = -\frac{1}{2\left(\Delta_{14} + \frac{|\Omega_c|^2}{\Delta_{12}^*}\right)}, \hspace{0.7cm}
    \rho_{43}^{(1)} = -\frac{1}{2\left(\Delta_{34} - \frac{|\Omega_c|^2}{\Delta_{23}^*}\right)}.
    \label{rho_eq}
\end{equation}  

The induced polarization $\vec{P} = \mathcal{N} \vec{d}\rho = \chi \vec{E}$ leads to the susceptibilities of the two circular components of the probe field, as follows.
\begin{equation}
\chi_+ = \frac{\mathcal{N}|d_{41}|^2}{\hbar}\rho_{41}^{(1)}, \hspace{0.7cm}
\chi_- = \frac{\mathcal{N}|d_{43}|^2}{\hbar}\rho_{43}^{(1)}\;.
\end{equation}  
In the presence of a magnetic field, the probe-induced coherences \(\rho_{41}\) and \(\rho_{43}\) differ, giving rise to unequal susceptibilities \(\chi_+\) and \(\chi_-\). The real part of the susceptibility determines the dispersion of the medium, and the difference \(\mathrm{Re}(\chi_- - \chi_+)\) produces birefringence, i.e., different refractive indices for the right- and left-circular polarizations. This birefringence manifests as a MOR angle of the probe polarization, quantified by \cite{RevModPhys.74.1153}  
\begin{equation}
    \theta(z) = \pi k_p L \, \mathrm{Re}\!\left(\chi_- - \chi_+\right),
    \label{theta_rho}
\end{equation}  
where \(k_p\) is the magnitude of the probe wave vector and L is the length of the medium.  
The imaginary part of the susceptibility, on the other hand, governs absorption. The difference \(\mathrm{Im}(\chi_- - \chi_+)\) thus corresponds to dichroism, which tends to reduce the transmitted probe intensity \cite{RevModPhys.74.1153}. 

\subsection{Atomic Response}

When a static magnetic field is applied to the atomic medium, the degeneracy of the magnetic sublevels is lifted, thereby inducing anisotropy in the system. This effect manifests in the real and imaginary parts of the atomic coherences $\rho_{41}$ and $\rho_{43}$ [see Eq.~(\ref{rho_eq})], corresponding to the left- and right-hand circularly polarized (LHC and RHC) probe transitions, respectively. Figures~\ref{rho41} and \ref{rho43} show the variation of these coherences as functions of the Zeeman splitting $B$. The real parts exhibit opposite dispersion profiles for the two circular components (the profiles interchange with $B\rightarrow -B$, while the imaginary parts, associated with absorption, display an EIT window around $B=0$. Within this transparency window, the dispersion exhibits a nearly linear region.

The width of the EIT window and the slope of the dispersion are strongly dependent on the control-field strength $\Omega_c$. To systematically analyze this dependence, we present contour plots of the real and imaginary parts of the coherences as functions of both $\Omega_c$ and the normalized Zeeman splitting $B/\gamma$ in Fig.~\ref{contour}.

As shown in Figs.~\ref{2a} and \ref{2b}, the real parts $\mathrm{Re}(\rho_{41})$ and $\mathrm{Re}(\rho_{43})$ clearly exhibit opposite dispersion profiles in the presence of the magnetic field, for all $\Omega_c$. This asymmetry produces a phase mismatch between the two circularly polarization components, resulting in polarization rotation of the probe beam, as described by Eq.~(\ref{theta_rho}). In contrast, the imaginary parts $\mathrm{Im}(\rho_{41})$ and $\mathrm{Im}(\rho_{43})$ display transparency windows (central blue regions in Figs.~\ref{2d} and \ref{2e}), which are characteristic signatures of EIT) generated by the control field. The width of the EIT window increases with increasing $\Omega_c$, indicating enhanced transparency.

Under these conditions, absorption is strongly suppressed, and the dichroism, proportional to $\mathrm{Im}(\rho_{43}-\rho_{41})$, remains nearly zero, as shown in Fig.~\ref{2f}. In contrast, birefringence quantified by the MOR angle and proportional to $\mathrm{Re}(\rho_{43}-\rho_{41})$ [see Eq.~(\ref{theta_rho})]---attains significant values for nonzero magnetic fields over a wide range of control field strengths. Consequently, the system enables substantial polarization rotation with minimal probe attenuation,
which is highly desirable for applications such as precision magnetometry.  

\begin{figure}
    \centering
    \begin{subfigure}{1\linewidth}
        \includegraphics[width=8cm]{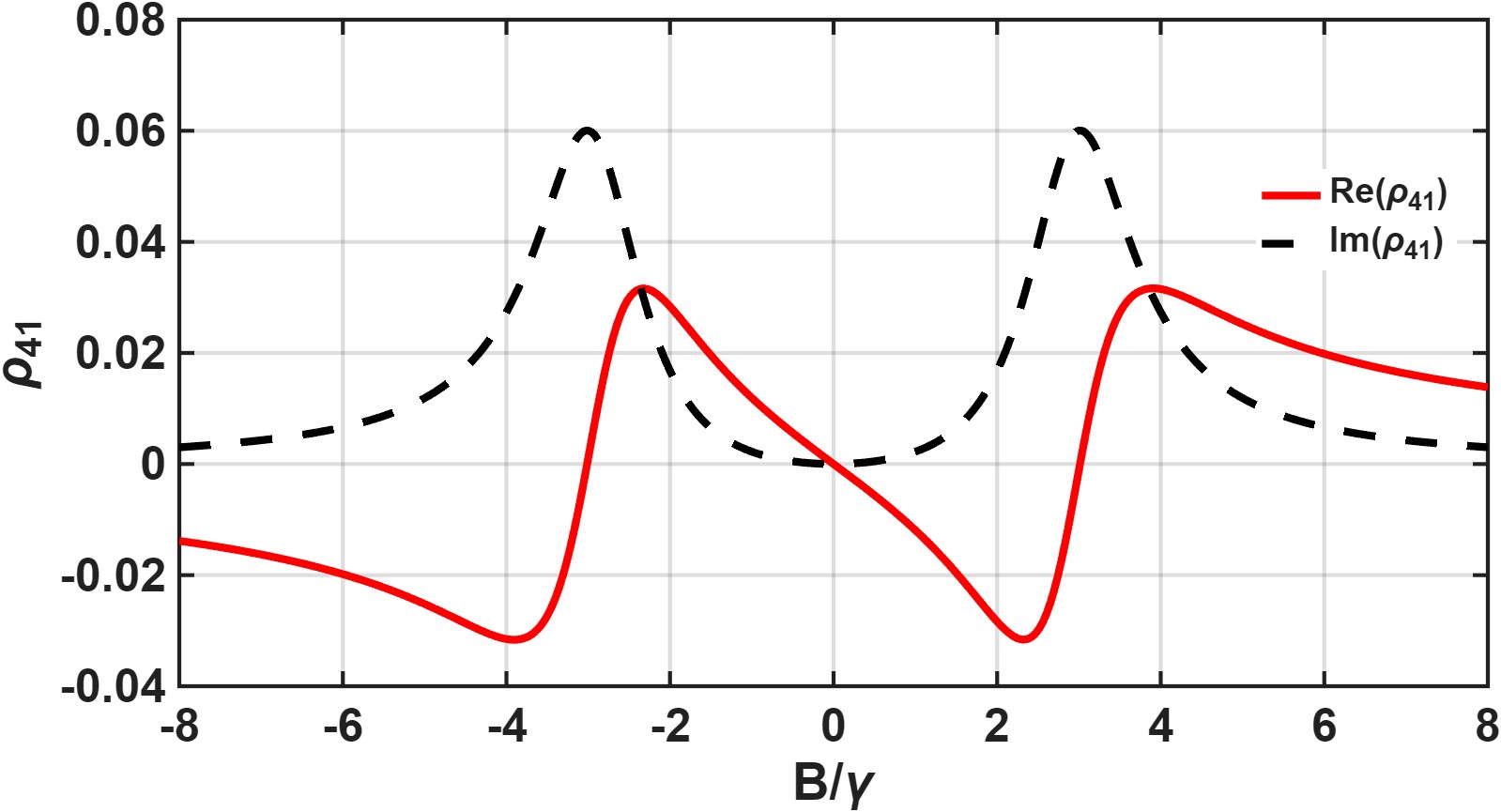}
    \caption{}
    \label{rho41}
    \end{subfigure}
    \hfill
    \begin{subfigure}{1\linewidth}
        \includegraphics[width=8cm]{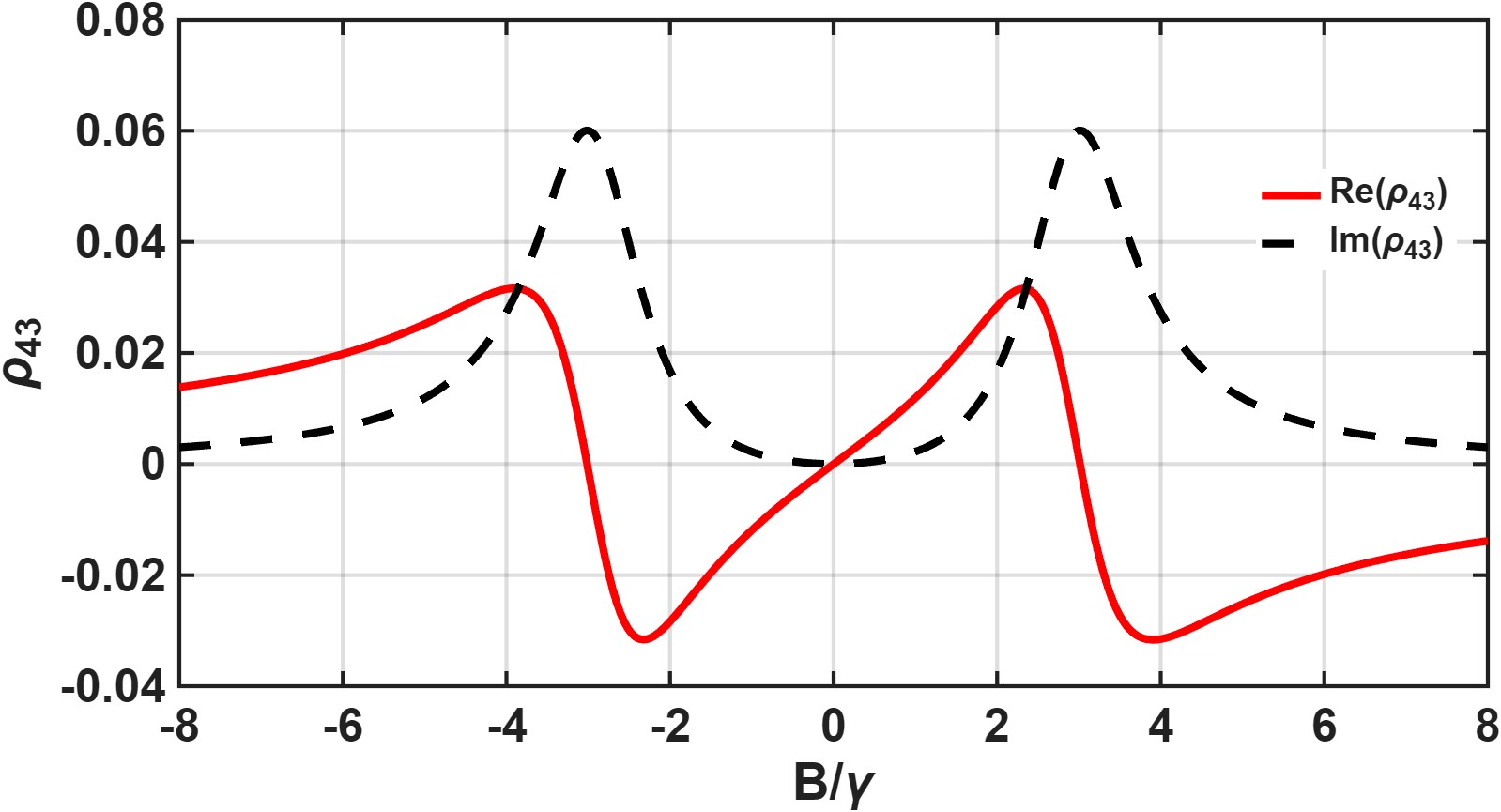}
    \caption{}
    \label{rho43}
    \end{subfigure}
   \caption{Variation of the atomic coherences $\rho_{41}$ and $\rho_{43}$ as functions of the normalized magnetic field $B/\gamma$. The results correspond to a control field strength $\Omega_c = 3\gamma$, equal probe components $\Omega_{+}=\Omega_{-}=0.2\gamma$, and resonance conditions $\Delta_c=\Delta_p=0$.}
    \label{rho}
\end{figure}

\begin{figure*}
\centering
\begin{subfigure}{.3\linewidth}
    \includegraphics[width=5.2cm]{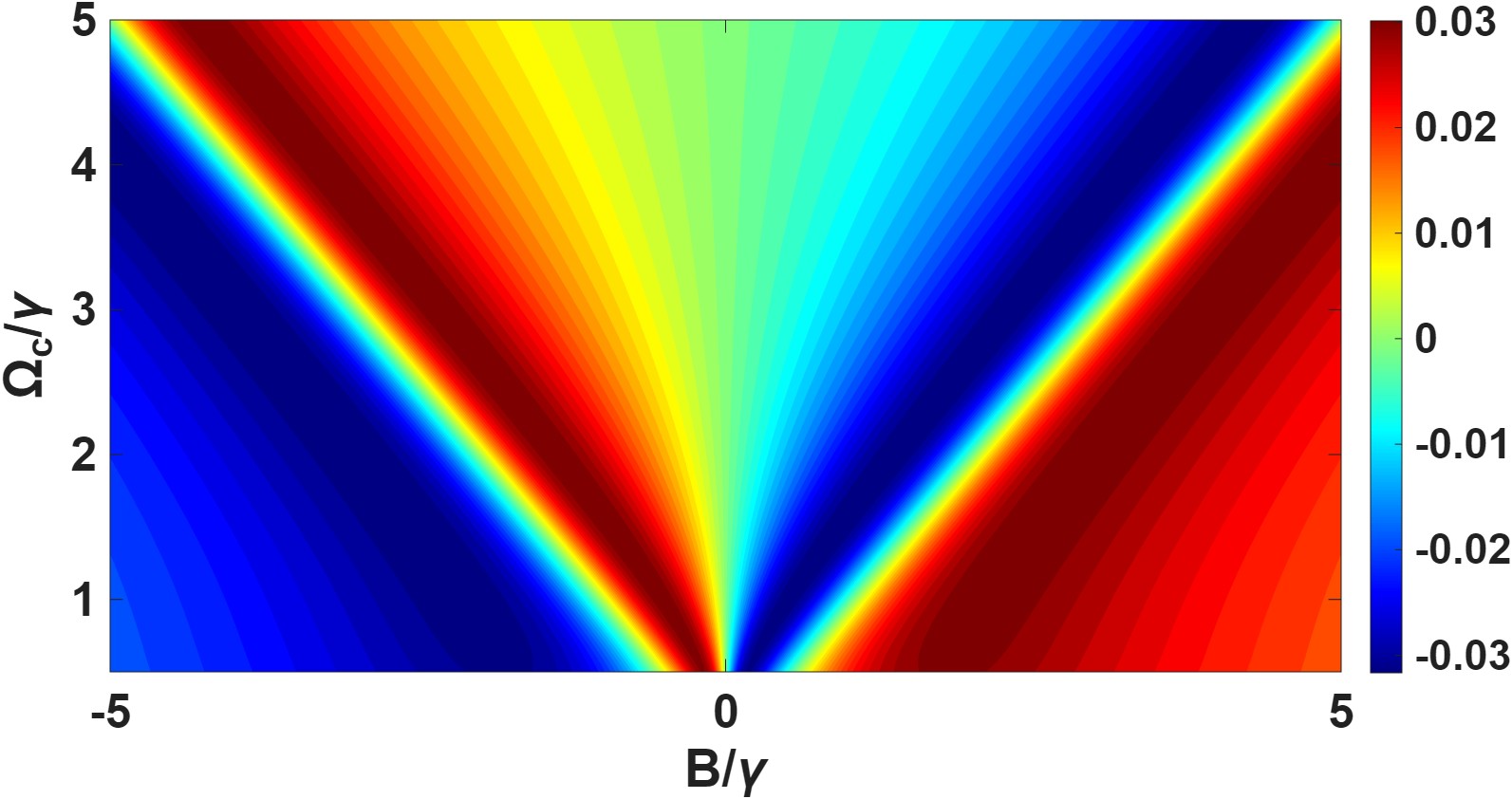}
    \caption{}
    \label{2a}
\end{subfigure}
\hfill
\begin{subfigure}{.3\linewidth}
    \includegraphics[width= 5.2cm]{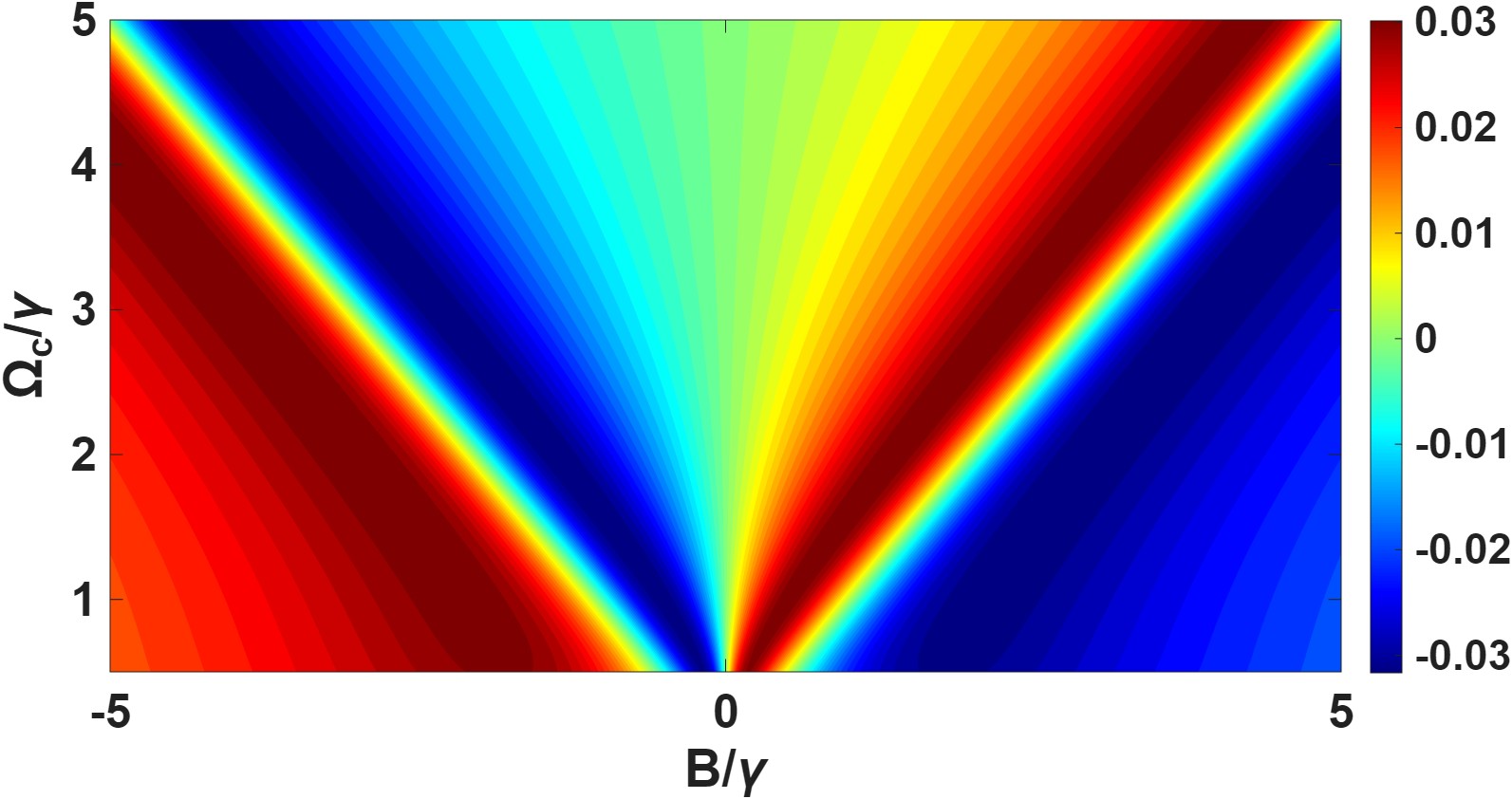}
    \caption{ }
    \label{2b}
\end{subfigure}
\hfill
\begin{subfigure}{.3\linewidth}
    \includegraphics[width=5.2cm]{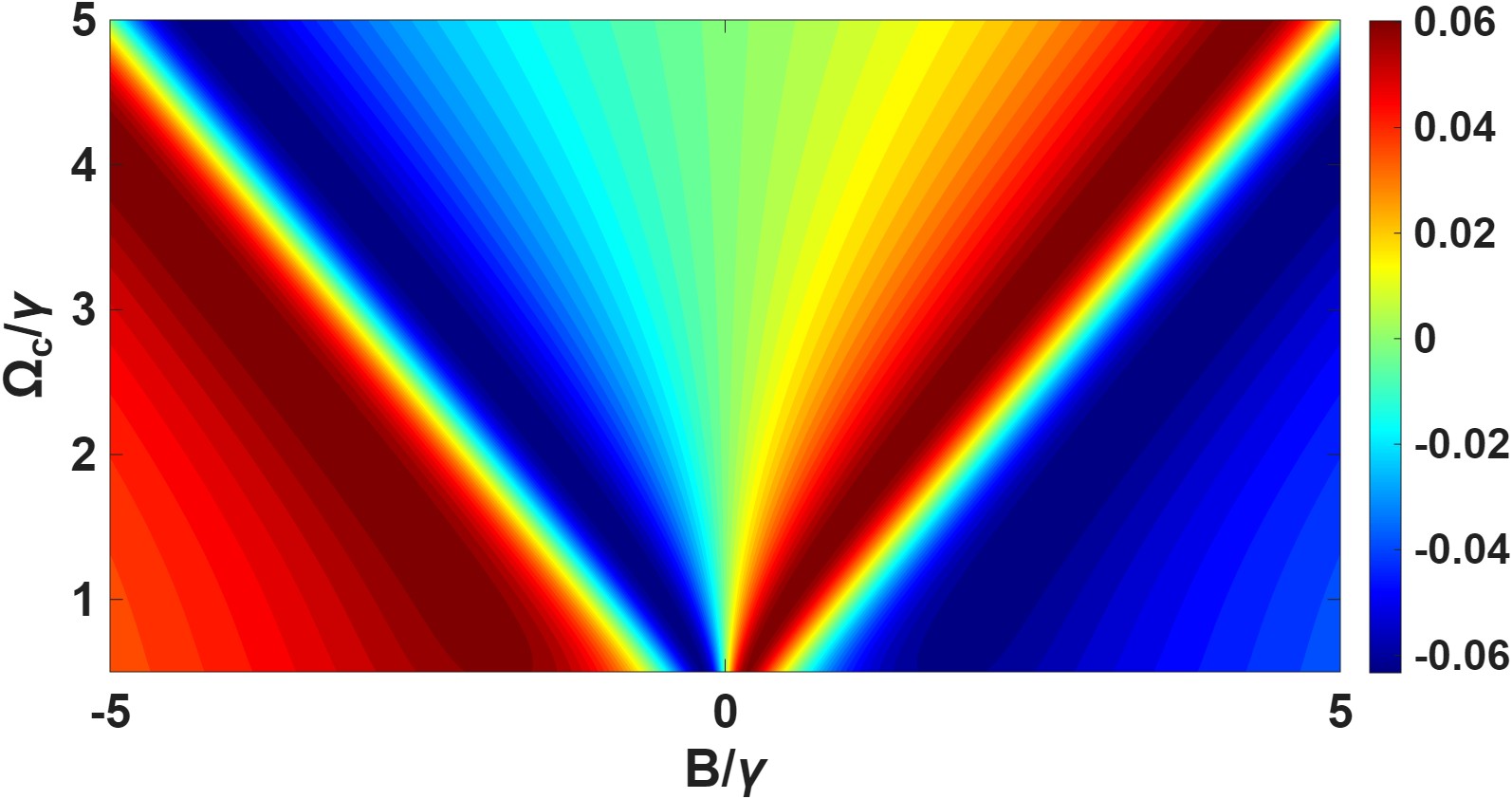}
    \caption{}
    \label{2c}
\end{subfigure}
\hfill
\begin{subfigure}{.3\linewidth}
    \includegraphics[width= 5.2cm]{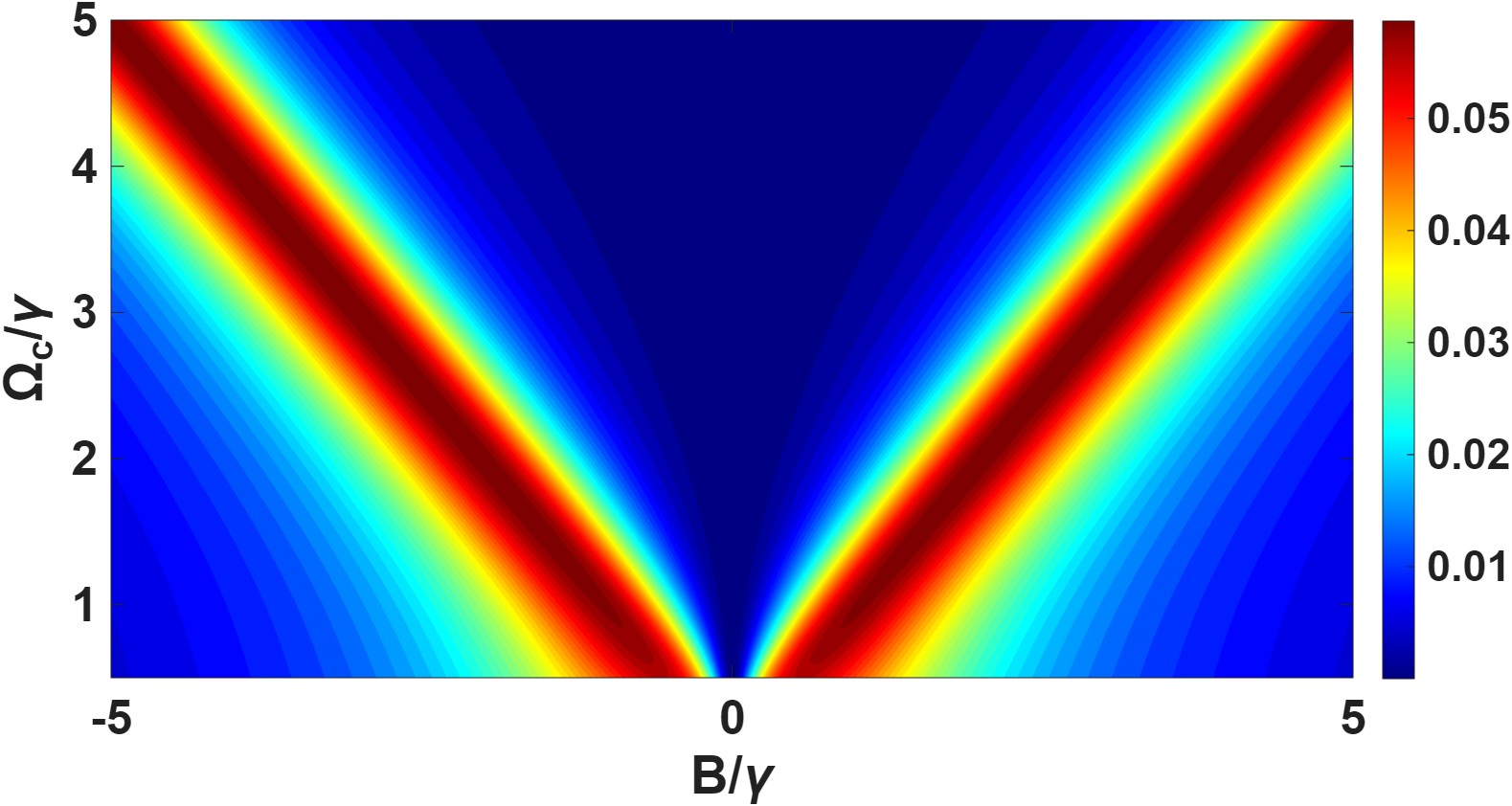}
    \caption{ }
    \label{2d}
\end{subfigure}
\hfill
\begin{subfigure}{.3\linewidth}
    \includegraphics[width=5.2cm]{imag_41.jpg}
    \caption{}
    \label{2e}
\end{subfigure}
\hfill
\begin{subfigure}{.3\linewidth}
    \includegraphics[width= 5.2cm]{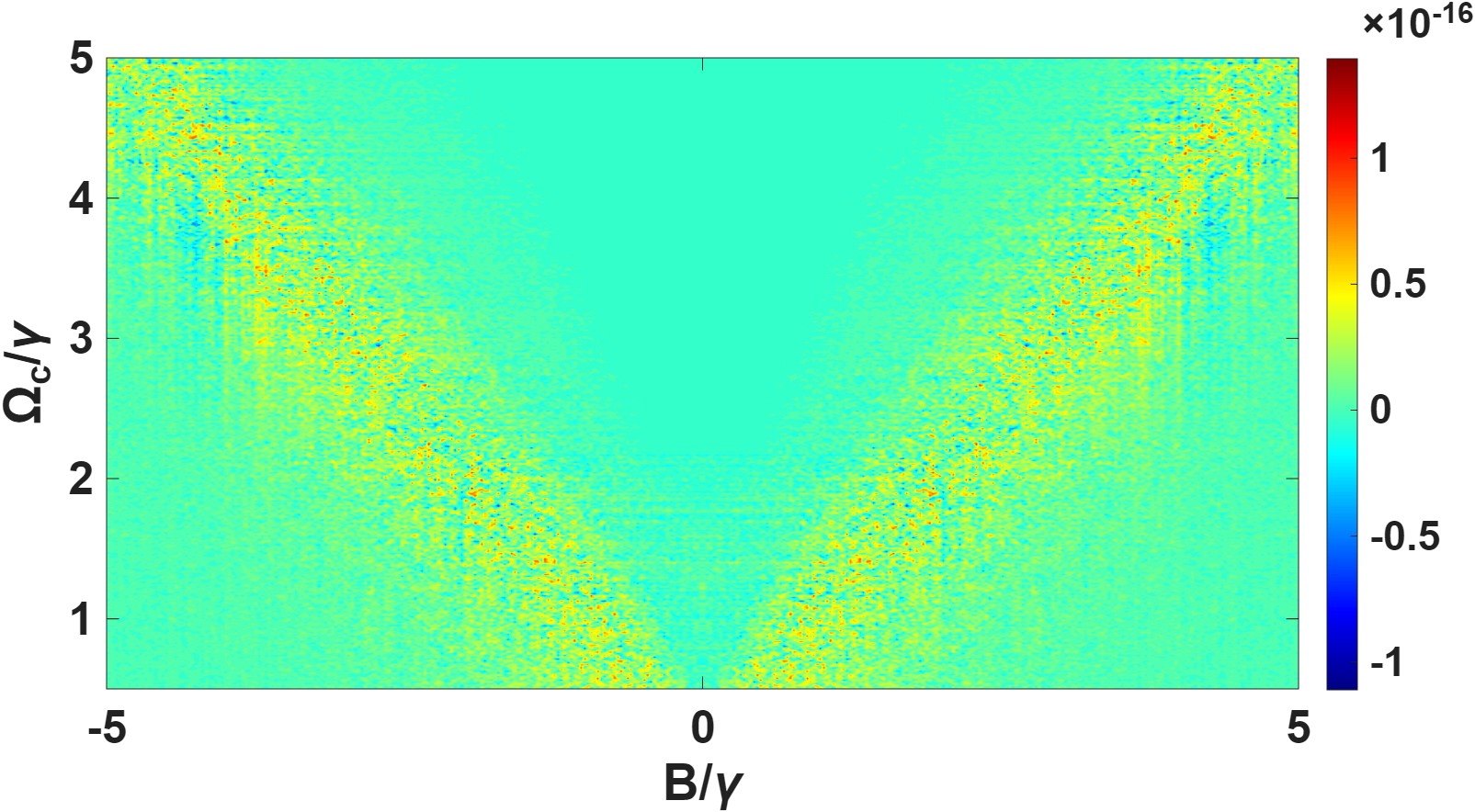}
    \caption{ }
    \label{2f}
\end{subfigure}
\hfill
\caption{Contour plots of the probe coherence as functions of the normalized control field strength $\Omega_c/\gamma$ and magnetic field $B/\gamma$. The color bars show the real parts in panels (a)--(c) and the imaginary parts in panels (d)--(f) of $\rho_{41}$, $\rho_{43}$, and $\rho_{43}-\rho_{41}$, respectively. The calculations are performed for equal probe components $\Omega_{+}=\Omega_{-}=0.2\gamma$ under resonance conditions $\Delta_c=\Delta_p=0$.}
 \label{contour}   
\end{figure*}

In the above discussion, the probe field was treated as a plane wave to illustrate how the atomic coherence varies with the applied magnetic field. To directly visualize these effects, we now consider a radially polarized LG beam, whose LHC and RHC polarization components acquire different phase shifts during propagation through the anisotropic medium.

\section{\label{iia} Detection of MOR Angle}

\begin{figure*}
\centering
\begin{subfigure}{.4\linewidth}
    \includegraphics[width=8cm]{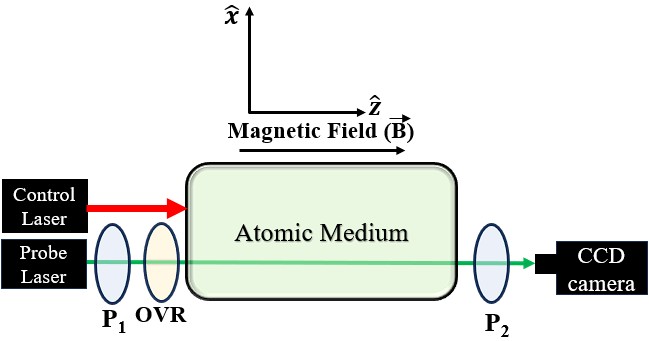}
    \caption{}
    \label{commonpath}
\end{subfigure}
\hfill
\centering
\begin{subfigure}{.5\linewidth}
    \includegraphics[width=10cm]{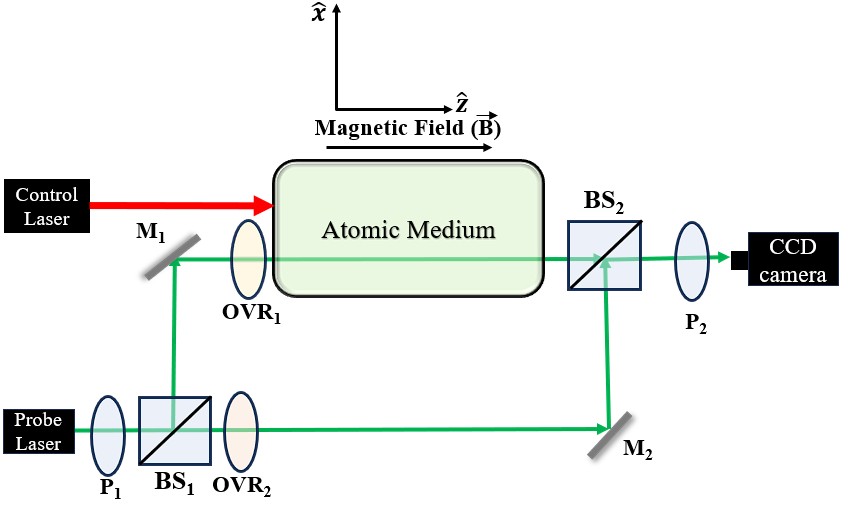}
    \caption{}
    \label{mzi}
\end{subfigure}
\hfill
\caption{ Schematic of the proposed (a) common-Path Interference based structured-light magnetometer and (b) Mach--Zehnder interferometric imaging magnetometer. OVR denotes an optical vortex retarder used to generate a radially polarized LG beam. Here, $P_1$ denotes an $x$-polarizer, while $P_2$ denotes an $x$-polarizer in Fig.~(a) and a circular polarizer in Fig.~(b). BS represents a beam splitter.}
\end{figure*}

To detect the MOR angle, we consider that a radially polarized LG probe field interacts with an ensemble of Rubidium atoms subjected to a homogeneous static magnetic field along the $z$ axis.  
In cylindrical coordinates, the electric field component of the  signal probe beam (with angular frequency $\omega$) at the input plane $z=0$ is \cite{PhysRevA.45.8185}
\begin{equation}
\begin{aligned}
\vec{E}_{p}(\rho,0,t)
 &= \mathcal{E}_
 {0p} \,\mathcal{L}^{\ell_1}_{m}(\rho) e^{i \omega t} e^{i \ell_1 \phi}\,\hat{\rho},\\
\mathcal{L}^{\ell_1}_{m}(\rho)
 &= e^{-\rho^2/w_0^2}\!\left(\frac{\sqrt{2}\rho}{w_0}\right)^{|\ell_1|}
    L_{m}^{|\ell_1|}\!\left(\tfrac{2\rho^2}{w_0^2}\right),
\end{aligned}
\end{equation}
where $\mathcal{E}_{0p}$ denotes the field amplitude, $L_m^{|\ell_1|}$ is the generalized Laguerre polynomial, $m$ represents the radial index corresponding to the number of concentric rings in the intensity distribution, and $\ell_1$ denotes the topological charge. The radial coordinate is given by $\rho=\sqrt{x^2+y^2}$, while $\phi=\tan^{-1}\!\left(\frac{y}{x}\right)$ defines the azimuthal angle, and \(w_0\) denotes the beam waist. 

The radially polarized LG mode can be expressed as a superposition of RHC and LHC polarized components:
\begin{equation}
\vec{E}_{p}=\vec{E}_{-}+\vec{E}_{+},
\end{equation}
with
\begin{equation}
\begin{aligned}
\vec{E}_{-} &= \mathcal{E}_{0-} \,\mathcal{L}^{\ell_1}_{m}(\rho)\,
e^{i(\ell_1+1)\phi}e^{-i\omega t}\hat{\sigma_-},\\
\vec{E}_{+} &= \mathcal{E}_{0+} \,\mathcal{L}^{\ell_1}_{m}(\rho)\,
e^{i(\ell_1-1)\phi}e^{i\omega t}\hat{\sigma_+}.
\end{aligned}
\end{equation}
where \(\sigma_+\) = \(\frac{1}{\sqrt{2}}(\hat{x}+i\hat {y})\), and \(\sigma_-\) = \(\frac{1}{\sqrt{2}}(\hat{x}-i\hat{y})\). Now the field propagation through the atomic medium is governed by the  Maxwell equations under the slowly varying envelope approximation in steady state \cite{Zienau_1975}
\begin{equation}
\frac{\partial E_{+}}{\partial z}=i2\pi k_p \chi_{+}(B)E_{+},\qquad
\frac{\partial E_{-}}{\partial z}=i2\pi k_p \chi_{-}(B)E_{-},
\end{equation}
whose solutions are
\begin{equation}
\begin{aligned}
E_{+}(z) &= E_{+}(0)\,
e^{-2\pi k_p L\operatorname{Im}\chi_{+}(B)}
e^{i\bigl[(\ell_1-1)\phi + 2\pi k_p L\operatorname{Re}\chi_{+} + k_pz_1\bigr]},\\
E_{-}(z) &= E_{-}(0)\,
e^{-2\pi k_pL\operatorname{Im}\chi_{-}(B) }
e^{i\bigl[(\ell_1+1)\phi + 2\pi k_p L\operatorname{Re}\chi_{-}(B) +k_pz_1\bigr]}.
\end{aligned}
\end{equation}

Here $L$ is the length of the atomic medium and \(z_1\) is rest of the  propagation distance of the beam through free space. In the limit of negligible absorption, 
$\operatorname{Im}\chi_{\pm}\approx0$, the signal probe-field intensity becomes
\begin{equation}
|E_{p}(z)|^{2}=|\mathcal{E}_{0p}|^{2}\bigl[\mathcal{L}^{\ell_1}_{m}(\rho)\bigr]^2,
\end{equation}
where $|\mathcal{E}_{0p}|^{2}=\mathcal{E}_{0+}^2 +\mathcal{E}_{0-}^2$.

Since the accumulated phase does not appear directly in the intensity distribution of the LG beam, an interference-based detection scheme is required to convert the phase information into a measurable intensity variation. We first consider a \emph{common-path interference} technique as the primary detection method, as illustrated in Fig.~\ref{commonpath}. In this configuration, the $\sigma^{+}$ and $\sigma^{-}$ polarization components of the same optical field are projected onto a common linear polarization basis and coherently superposed. Their relative phase, which contains the magnetic-field-dependent magneto-optical rotation, is thereby converted into a spatial intensity modulation. As an extension of this approach, we consider a Mach--Zehnder interferometer (MZI)-based configuration, shown in Fig.~\ref{mzi}, in which a separate reference beam is introduced to interfere with the signal field. The two detection schemes are discussed separately in the following sections.

\subsubsection{\label{comInt} Common-Path Interference based Structured-Light Magnetometer}

To measure the MOR angle induced by the RHC and LHC components in the presence of a magnetic field, the transmitted field is passed through a polarizer P$_2$ which is oriented along $x$-polarization direction. The resulting electric field is given by
\begin{equation}
E_x=\frac{1}{\sqrt{2}}\mathcal{E}_{0p}e^{i(\ell_1\phi+k_pz_1)}
\left[
e^{-i(\phi-\delta_+(B))}
+
e^{i(\phi+\delta_-(B))}
\right]\;,
\end{equation}
where \(\delta_{\pm} = 2\pi k_pL\mathrm{Re}\chi_{\pm}\). As shown in Figs.~\ref{2a} and \ref{2b}, the real parts of the susceptibilities satisfy
\(\mathrm{Re}(\chi_+)=-\mathrm{Re}(\chi_-)\).
Using this relation, the output intensity after \(P_2\) becomes
\begin{equation}
I_x(B)=
|\mathcal{E}_{0p}|^2
|\mathcal{L}^{\ell_1}_{m}|^2
\left[
1+\cos\left(2\left(\phi+\theta(B)\right)\right)
\right],
\label{intenstyx}
\end{equation}
where the MOR angle is defined as \(\theta(B)
=
\pi k_pL
\left[
\mathrm{Re}(\chi_-)-\mathrm{Re}(\chi_+)
\right]
= \frac{1}{2}(\delta_-(B)-\delta_+(B))=
2\pi k_pL\,\mathrm{Re}(\chi_-)\).

It is evident from Eq.~(\ref{intenstyx}) that the azimuthal modulation term,
$\cos(2(\phi+\theta(B)))$, is independent of the OAM
index $\ell_1$. Therefore, irrespective of the OAM value, this scheme always produces a two-petal intensity pattern. The OAM index influences only the radial envelope through the LG mode
$|\mathcal{L}^{\ell_1}_{m}|^2$. As $|\ell_1|$ increases, the point of the intensity maximum shifts outward and its distance from the center changes according to \(\rho_{\mathrm{max}}
=
w_0
\sqrt{\frac{|\ell_1|}{2}}\),
thereby increasing the separation between the two petals without changing their number.

The magnetic-field-induced MOR angle appears as a phase shift in the azimuthal dependence, resulting in a rotation of the two-petal intensity pattern in the transverse plane. The intensity is plotted as $|\Omega_p|^2$, and is therefore expressed in units of $\gamma$. Figures~\ref{c1}--\ref{c4} show the transverse intensity distribution $I_x$ for different medium lengths $L$ at a fixed magnetic-field strength of $B=0.5\gamma$. The figures clearly demonstrate the rotation of the petal-shaped intensity pattern. As the length of the medium increases, the rotation angle of the beam also increases, in agreement with the results presented in Fig.~\ref{com}. This behavior arises from the larger phase accumulation experienced by the beam as it propagates through a longer interaction region. However, the length of the medium must remain smaller than the Rayleigh length, \(z_R=\pi w_0^2/\lambda\), to ensure that diffraction effects remain negligible and that the transverse beam profile is preserved throughout the interaction region \cite{Paschotta__rayleigh_length}. Therefore, in the subsequent analysis, we fix the medium length at \(L=50~\mathrm{mm}\), which is smaller than the corresponding value of \(z_R\) for the chosen wavelength.
Figures~\ref{c5}--\ref{c8} illustrate the influence of the magnetic field \(B\) on the transverse intensity distribution. As the magnetic field strength is increased from \(0\) to \(1.2\gamma\), the beam pattern exhibits a continuous rotation. This behavior is a direct consequence of the increase in the magneto-optical rotation angle induced by the applied magnetic field. Throughout this work, the magnetic field strength \(B\) is expressed in units of \(\gamma\).

This rotation angle can be quantified by determining the angular position of the intensity maxima. The condition for maximum intensity is 
\(2(\phi+\theta)=2n\pi\) (from Eq.\ref{intenstyx},)
where $n$ is an integer. Consequently, the equation describing the orientation of the petals in the transverse plane is
\begin{equation}
y=x\tan\left(n\pi-\theta\right).
\end{equation}
Thus, the MOR angle can be obtained directly by monitoring the rotation of the petal orientation.

Alternatively, the MOR angle can be extracted from the intensity measured at a
fixed point $(x,y)$ in the transverse plane. Since the local intensity changes
continuously with the rotation of the petal pattern, from the Eq.\ref{intenstyx} the MOR angle can be expressed as

\begin{equation}
\theta
=
\frac{1}{2}\cos^{-1}
\left(
\frac{I_x(B)}{I_0}-1
\right)- \phi,
\end{equation}
where \(I_0=
|\mathcal{E}_{0p}|^2
|\mathcal{L}^{\ell_1}_{m}|^2\).
Therefore, the MOR angle can be determined either by tracking the rotation of the two-petal pattern using an imaging detector or by monitoring the intensity variation at a fixed point in the transverse plane.

 However, this configuration is restricted to the generation of only a two-petal beam pattern. A tunable multi-petal structure can be realized by interfering the beam with another radially polarized beam possessing a different OAM index. The implementation and analysis of this MZI -based scheme are discussed in the next section.

\begin{figure*}
\centering
\begin{subfigure}{.2\linewidth}
    \includegraphics[width=4cm]{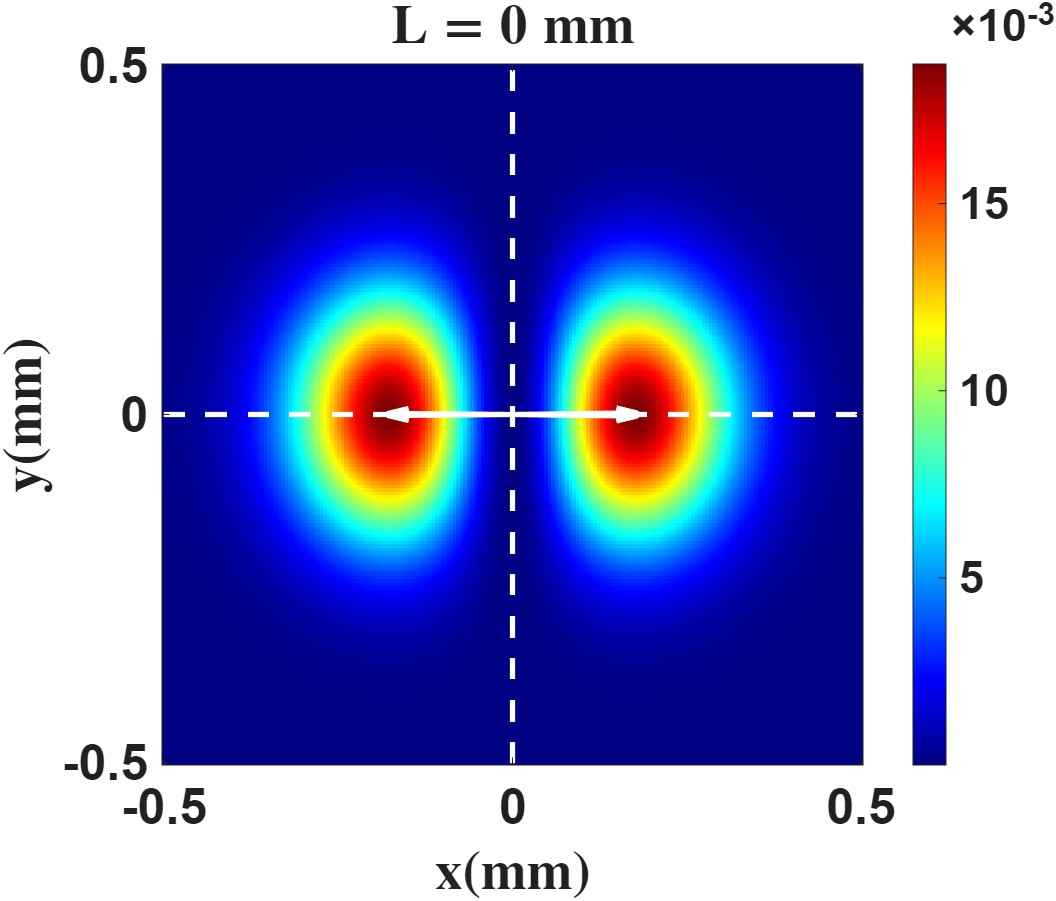}
    \caption{}
    \label{c1}
\end{subfigure}
\hfill
\begin{subfigure}{.2\linewidth}
    \includegraphics[width= 4cm]{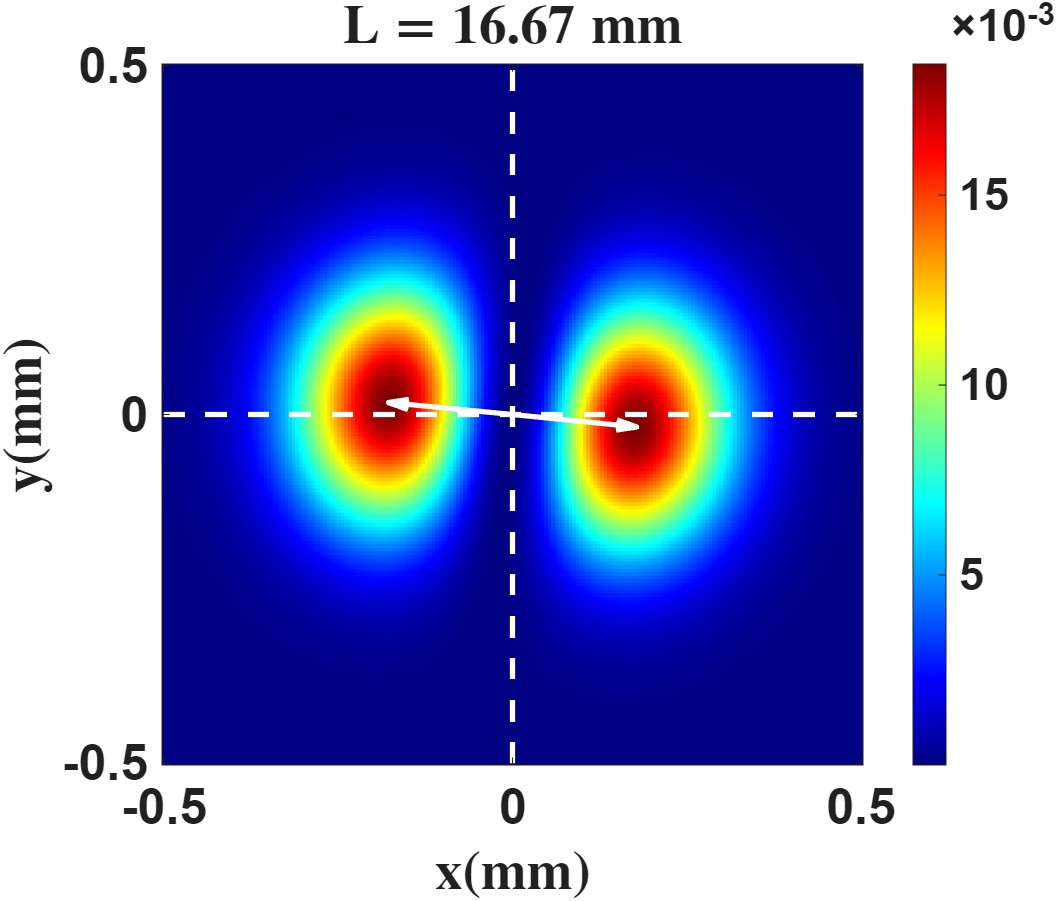}
    \caption{ }
    \label{c2}
\end{subfigure}
\hfill
\begin{subfigure}{.2\linewidth}
    \includegraphics[width=4cm]{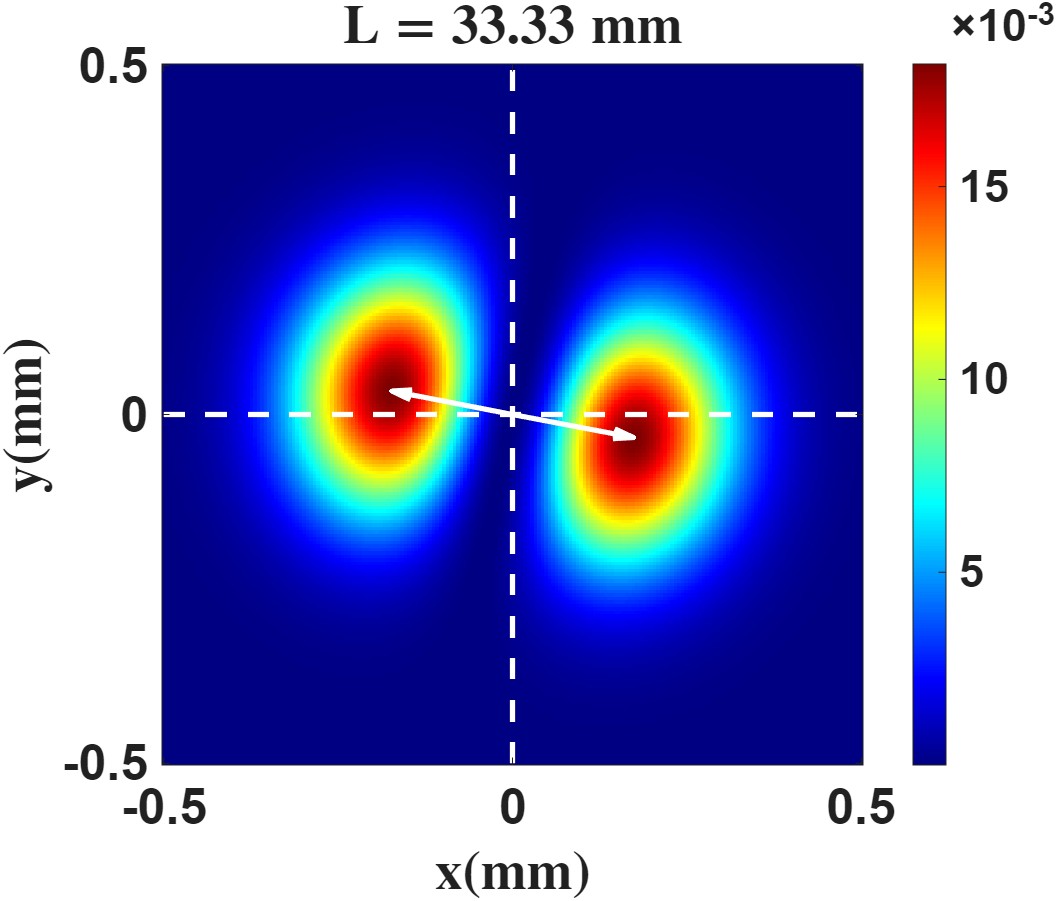}
    \caption{}
    \label{c3}
\end{subfigure}
\hfill
\begin{subfigure}{.2\linewidth}
    \includegraphics[width= 4cm]{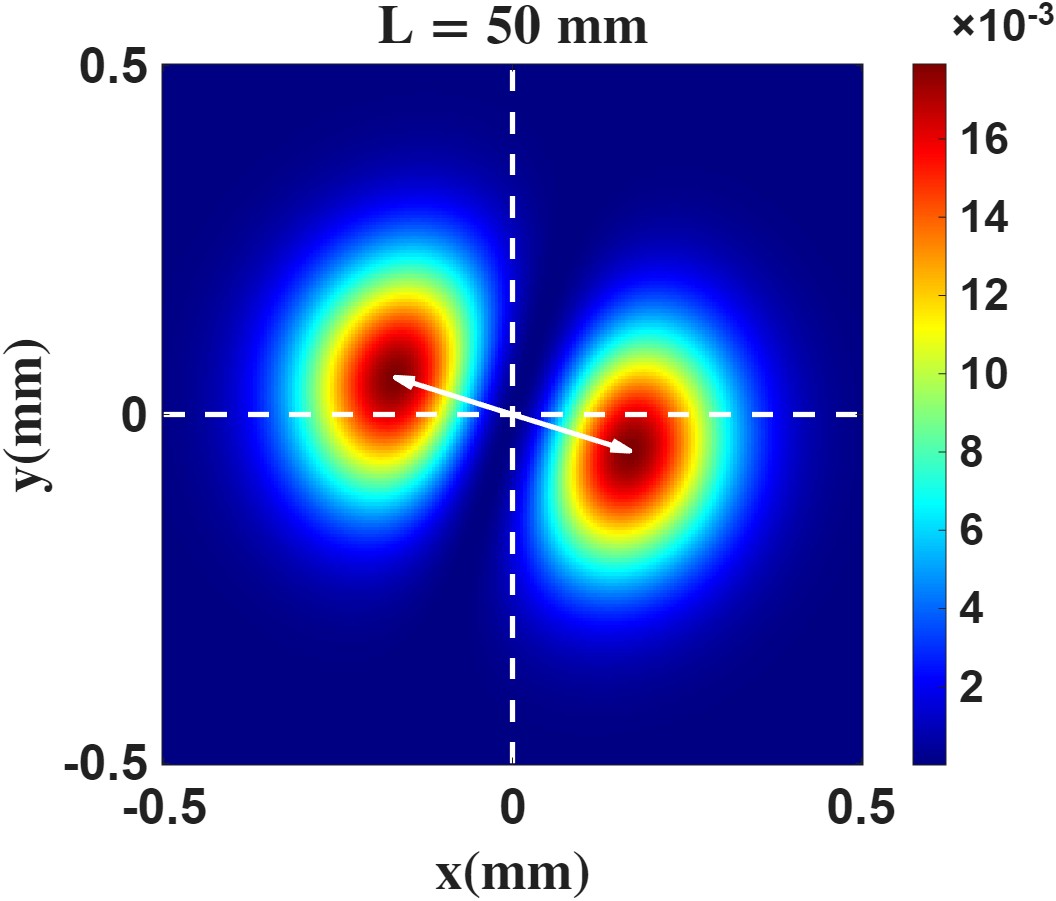}
    \caption{ }
    \label{c4}
\end{subfigure}
\hfill
\centering
\begin{subfigure}{.2\linewidth}
    \includegraphics[width=4cm]{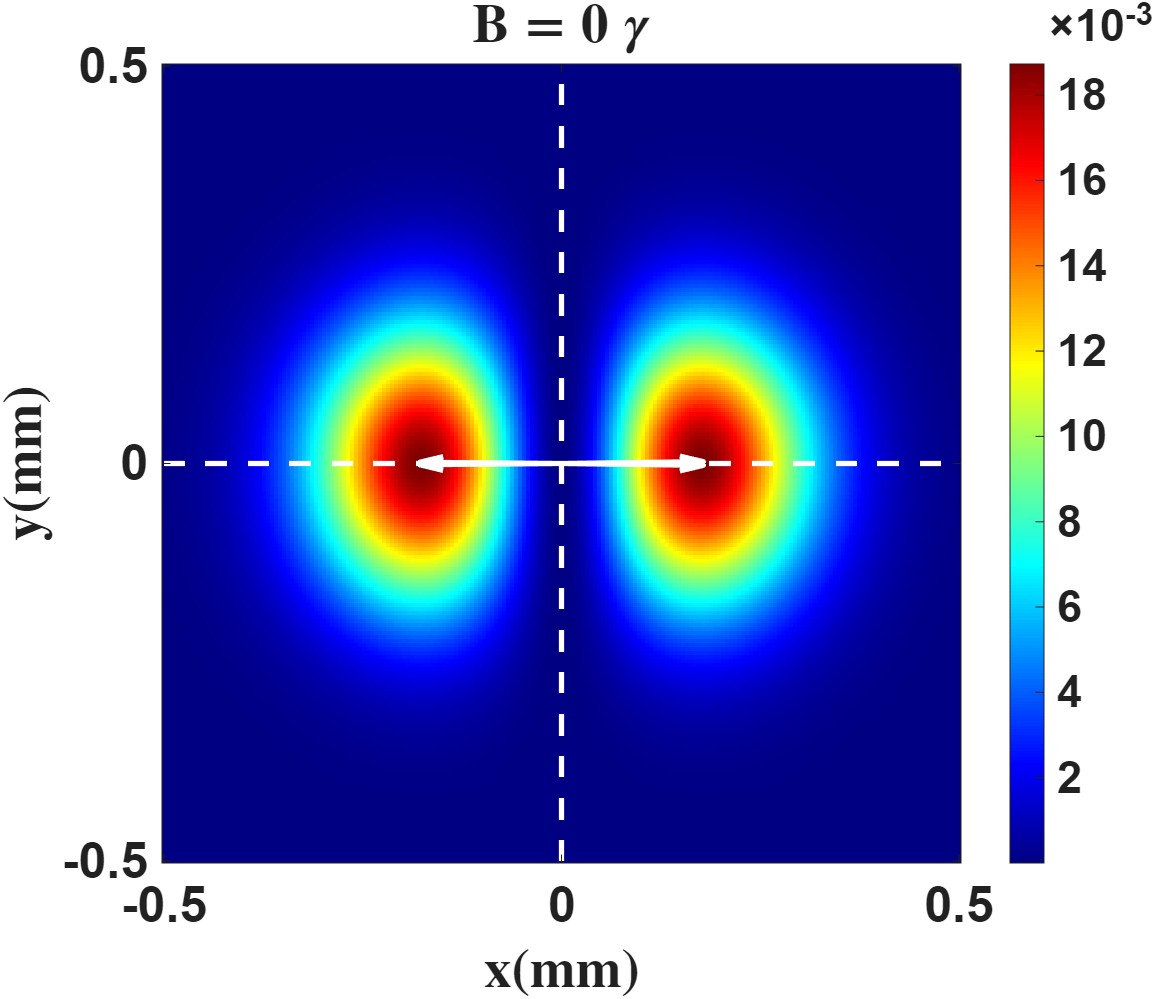}
    \caption{}
    \label{c5}
\end{subfigure}
\hfill
\begin{subfigure}{.2\linewidth}
    \includegraphics[width= 4cm]{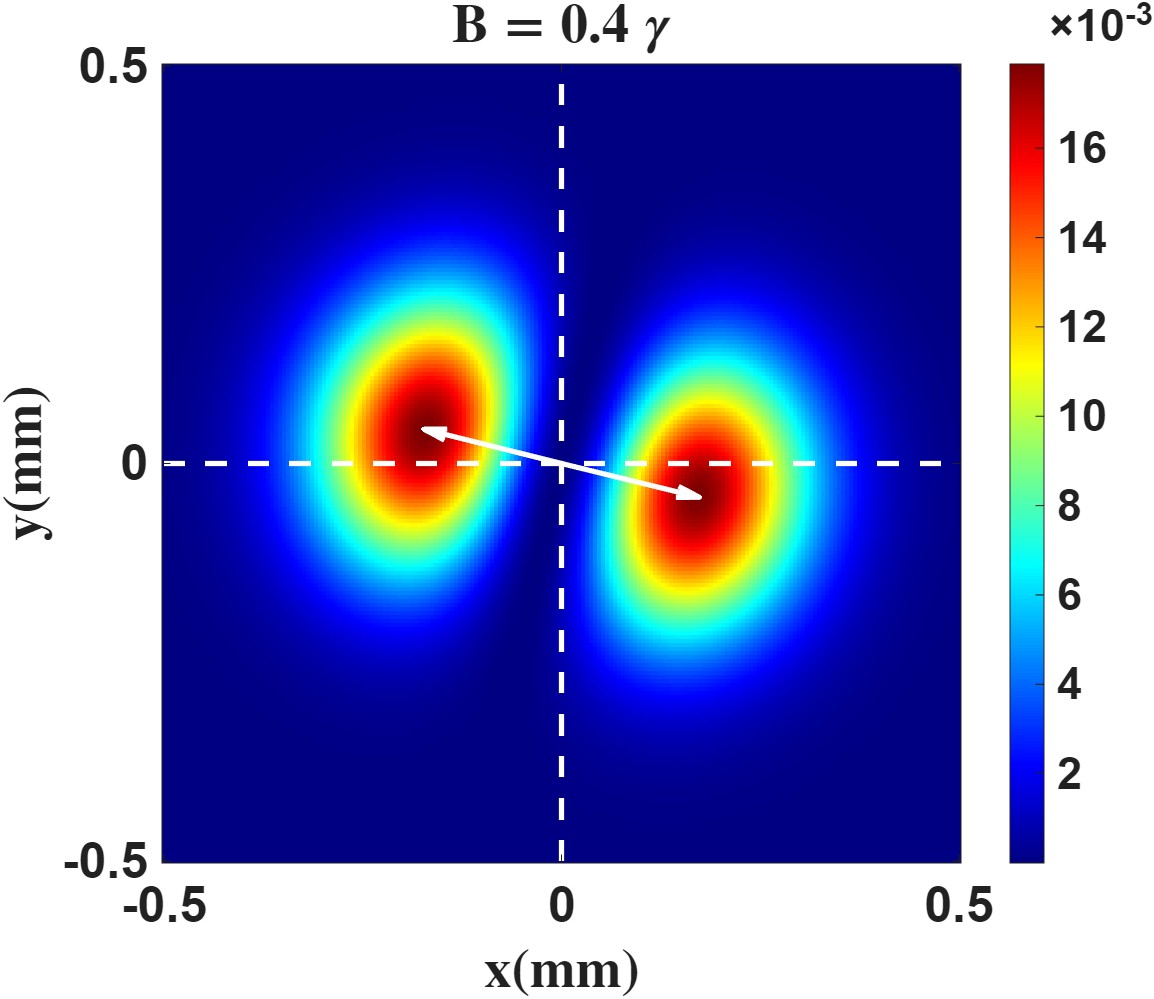}
    \caption{ }
    \label{c6}
\end{subfigure}
\hfill
\begin{subfigure}{.2\linewidth}
    \includegraphics[width=4cm]{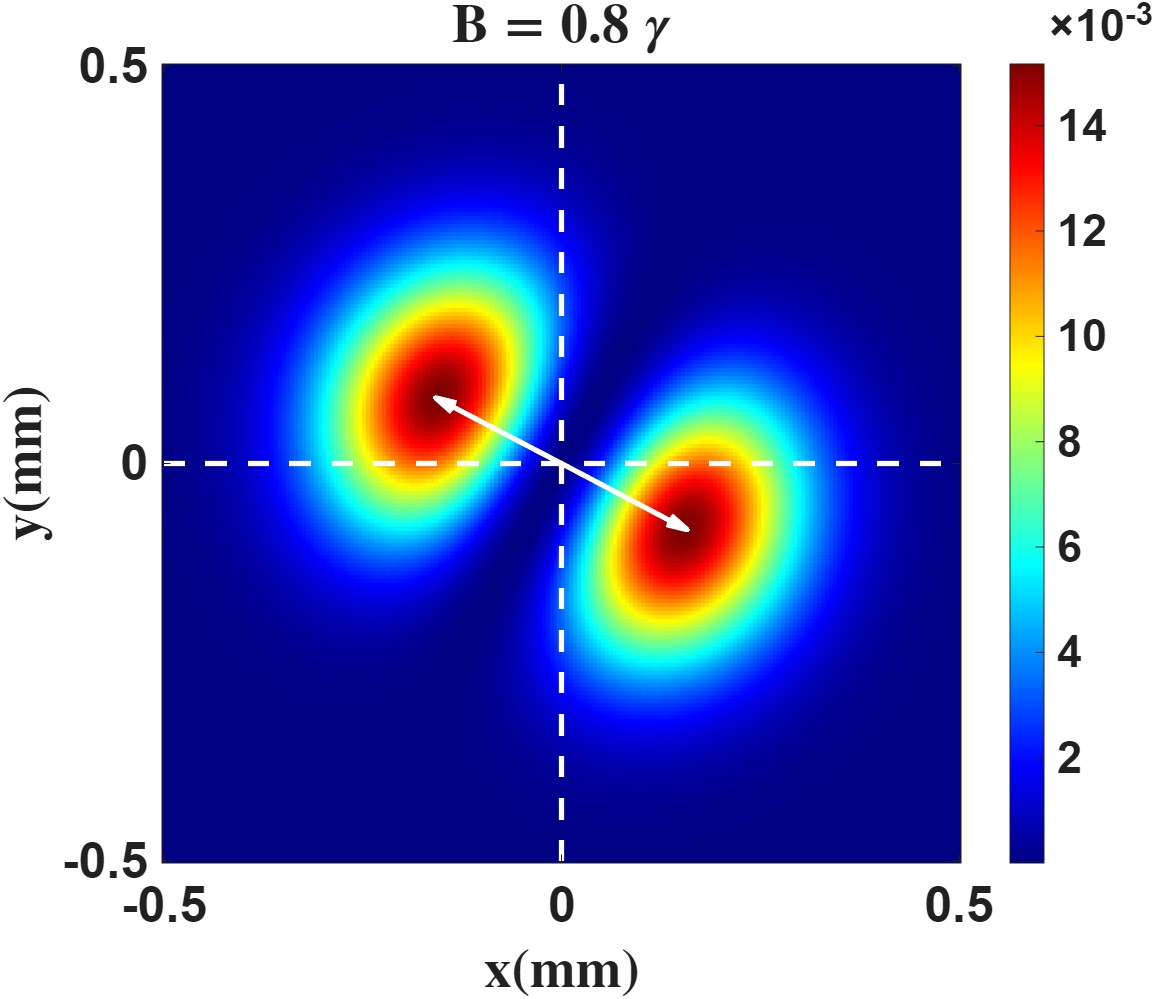}
    \caption{}
    \label{c7}
\end{subfigure}
\hfill
\begin{subfigure}{.2\linewidth}
    \includegraphics[width= 4cm]{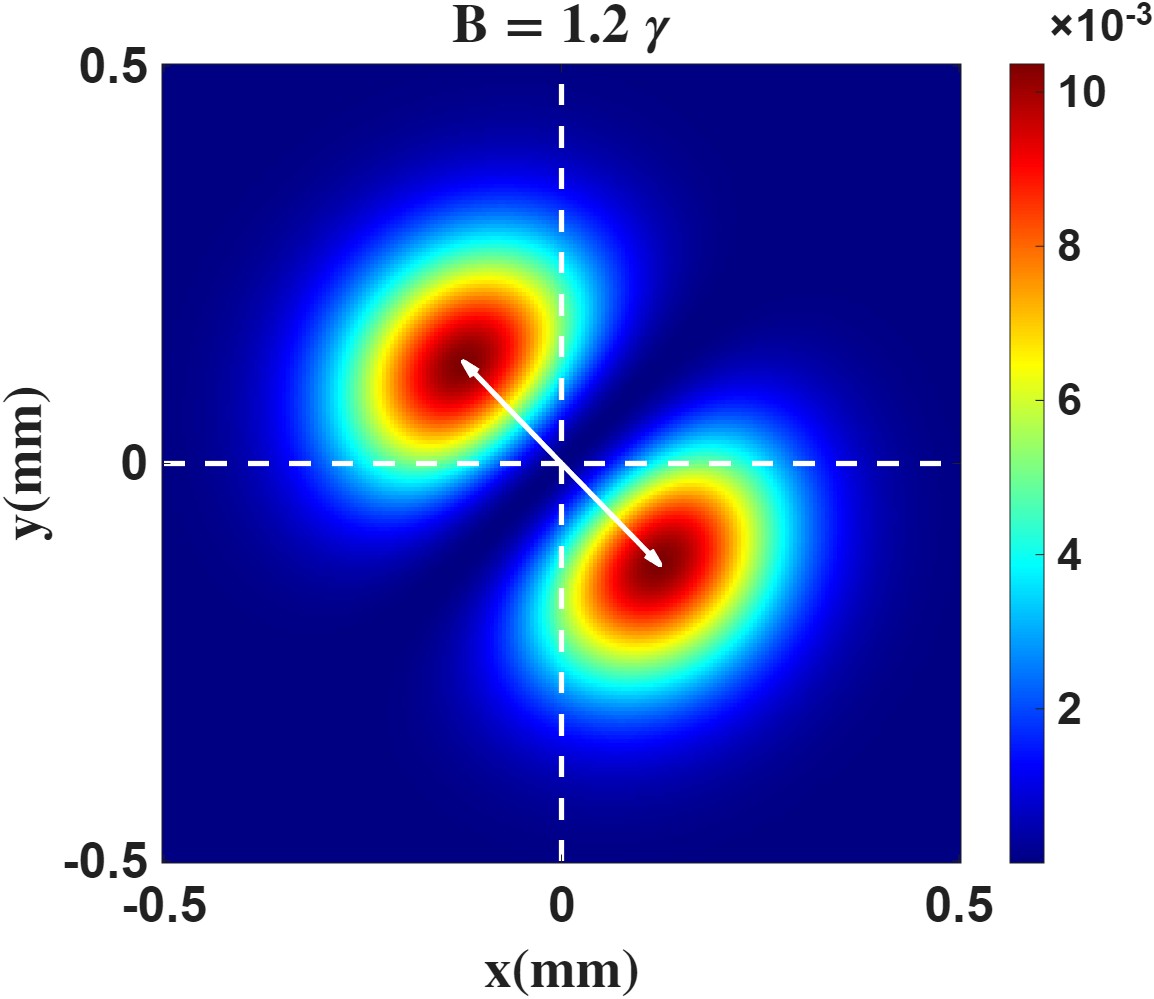}
    \caption{ }
    \label{c8}
\end{subfigure}
\hfill
\caption{Intensity distribution $I_x$ for (a)--(d) different propagation lengths of the atomic medium in the presence of a static magnetic field $B=0.5\gamma$, and (e)--(h) different magnetic-field strengths for a fixed medium length of $L=50~\mathrm{mm}$. The parameters are $\ell_1=1$, $\Omega_{0p}=0.2\gamma$, $\Omega_c=2.5\gamma$, $\Delta_p=\Delta_c=0$, atomic number density $N=2\times10^9~\mathrm{cm}^{-3}$, wavelength $\lambda=780~\mathrm{nm}$, radial index $m=0$, and beam waist $w_0=250~\mu\mathrm{m}$.}
 \label{com}
\end{figure*}

\subsubsection{Mach--Zehnder Interferometric Imaging Magnetometer}

An alternative approach for measuring the phase accumulated by the probe beam in the presence of a magnetic field is based on interferometry. In this scheme, the output probe beam is combined with a reference radially polarized LG beam in a Mach--Zehnder interferometer as shown in Fig. \ref{mzi}. The reference field is expressed as
\begin{equation}
\vec{E}_{\mathrm{ref}}
=
\mathcal{E}_{0\mathrm{ref}}
\mathcal{L}^{\ell_2}_{m}(\rho)
e^{ik_pz_2}
\left[
e^{i(\ell_2-1)\phi}\hat{\sigma}_{+}
+
e^{i(\ell_2+1)\phi}\hat{\sigma}_{-}
\right],
\end{equation}

with $z_2$ denoting the propagation distance of the reference beam. After passing through the interferometer, the signal and reference beams are projected onto the circular polarization basis using an appropriate polarizer \(P_2\). Assuming that the optical path difference satisfies $z_1-z_2\approx0$, the interference intensity corresponding to the $\sigma_{+}$ component is

\begin{equation}
I_{+}
=
\left|
\vec{E}_{s+}
+
\vec{E}_{\mathrm{ref}+}
\right|^{2}.
\end{equation}

Assuming equal amplitudes for the signal and reference beams,
$\mathcal{E}_{0\mathrm{ref}}=\mathcal{E}_{0p}$, the above expression reduces to

\begin{equation}
\begin{aligned}
I_+(B)
&=
\frac{1}{4}
|\mathcal{E}_{0p}|^2
\Big[
|\mathcal{L}^{\ell_1}_{m}(\rho)|^2
+
|\mathcal{L}^{\ell_2}_{m}(\rho)|^2
\\
&
\qquad
+
2
\mathcal{L}^{\ell_1}_{m}(\rho)
\mathcal{L}^{\ell_2}_{m}(\rho)
\cos\!\left(
(\ell_1-\ell_2)\phi
+\delta_+(B)
\right)
\Big].
\label{I+}
\end{aligned}
\end{equation}

Similarly, the interference intensity for the $\sigma_{-}$ component is

\begin{equation}
\begin{aligned}
I_-(B)
&=
\frac{1}{4}
|\mathcal{E}_{0p}|^2
\Big[
|\mathcal{L}^{\ell_1}_{m}(\rho)|^2
+
|\mathcal{L}^{\ell_2}_{m}(\rho)|^2
\\
&
\qquad
+
2
\mathcal{L}^{\ell_1}_{m}(\rho)
\mathcal{L}^{\ell_2}_{m}(\rho)
\cos\!\left(
(\ell_1-\ell_2)\phi
+\delta_-(B)
\right)
\Big].
\label{I-}
\end{aligned}
\end{equation}

The azimuthal modulation in Eqs.~(\ref{I+}) and (\ref{I-}) originates from the interference term
$\cos[(\ell_1-\ell_2)\phi]$, which results from the superposition of two LG beams carrying different OAM indices. Consequently, the transverse intensity exhibits a petal-like structure containing $|\ell_1-\ell_2|$ equally spaced bright lobes. The radial envelope of the interference pattern is determined by the product of the corresponding LG mode functions,
$\mathcal{L}^{\ell_1}_{m}(\rho)$ and
$\mathcal{L}^{\ell_2}_{m}(\rho)$.

In the absence of an external magnetic field, the phase offsets
\(
\delta_{\pm}(B)
\) vanish, and the interference pattern remains stationary in the transverse plane. When a longitudinal magnetic field is applied, magneto-optical rotation induces different phase shifts for the $\sigma_{+}$ and $\sigma_{-}$ components. These phase shifts appear as additional phase terms in the interference fringes,
\(
\cos\!\left[
(\ell_1-\ell_2)\phi+\delta_{\pm}
\right],
\)
causing the entire petal pattern to rotate. Therefore, the rotation angle of the interference pattern provides a direct measure of the accumulated phase offset, while the magneto-optical rotation angle is obtained from
\(
\theta
=
\frac{1}{2}(\delta_{-}
-
\delta_{+}).
\)
From the Eqs.(\ref{I+}) and 
(\ref{I-}), the angular position of the intensity maxima satisfies
\begin{equation}
y
=
x
\tan
\left[
\frac{2n\pi-\delta_{\pm}(z,B)}
{\ell_1-\ell_2}
\right],
\label{15}
\end{equation}
where $n$ is an integer. Thus, the phase offsets $\delta_{\pm}$ can be determined by tracking the rotation of any bright lobe in the transverse plane, and the MOR angle follows directly from their difference. Figures~\ref{sp1}--\ref{sp4} show the interference patterns corresponding to $|\ell_1-\ell_2|=3$ for \(I_+\)  and Fig. \ref{sp5}-\ref{sp8} for \(I_-\). As the magnetic field increases, the three-petal pattern associated with the $\sigma_{+}$ component rotates anticlockwise, whereas the $\sigma_{-}$ pattern rotates clockwise.

Alternatively, instead of tracking the angular displacement of the petals, the accumulated phase can be extracted from the intensity measured at a fixed point in the transverse plane. For the special case $\ell_1=-\ell_2=\ell$, Eqs.~(\ref{I+}) and (\ref{I-}) yields
\begin{equation}
\delta_{\pm}
=
\cos^{-1}
\left(
\frac{I_{\pm}(\rho,B)}
{I_{0\pm}(\rho)}
-1
\right)
-
2\ell\phi,
\label{16}
\end{equation}
where \(
I_{0\pm}(\rho)
=
\frac{1}{2}
|\mathcal{E}_{0p}|^2
|\mathcal{L}^{\ell}_{m}(\rho)|^2.
\)

Therefore, by monitoring the intensity at a fixed position in the transverse plane, the phase shifts \(\delta_{\pm}\), and consequently the MOR angle, can be determined without explicitly measuring the rotation angle of the interference pattern. Figure~\ref{IvsB} shows the variation of the intensity \(I_{+}\) as a function of the magnetic field strength \(B\) at the fixed position \((x,y)=(w_0/\sqrt{2},0)\). As the magnetic field strength increases, the measured intensity decreases owing to the spatial rotation of the petal-shaped interference pattern

The observed variation in intensity also depends on the interaction length of the medium, as discussed previously. In particular, for \(L=50~\mathrm{mm}\), the intensity exhibits a more rapid variation with the magnetic field, reflecting the enhanced phase accumulation over the longer propagation distance. To quantify this behavior further, we also evaluate the derivative of the intensity with respect to the magnetic field, \(dI_{+}/dB\), as shown in Fig.~\ref{slopvar}. Since this quantity is directly related to the sensitivity of the proposed magnetometer, its implications are discussed in greater detail in the next section.

An important feature of the Mach--Zehnder interferometric scheme is that the OAM indices of the signal and reference beams can be chosen independently, thereby providing additional control over the spatial structure of the interference pattern. Figure~\ref{ints} shows the intensity distribution \(I_{+}\) for \(B=1.2\gamma\). The results presented in Figs.~\ref{int4}--\ref{int7} correspond to the OAM combinations \((\ell_1,\ell_2)=(2,-2)\), \((3,-2)\), \((3,-3)\), and \((4,-3)\), respectively. Unlike the configuration depicted in Fig.~\ref{commonpath}, which is restricted to a two-petal structure, the present arrangement yields an interference pattern with \(|\ell_1-\ell_2|\) petals. Increasing $|\ell_1-\ell_2|$ increases the number of petals, and although the angular rotation of each individual petal,
\(
\Delta\phi
=
\frac{\delta_{\pm}}
{|\ell_1-\ell_2|},
\)
becomes smaller, the petals become progressively narrower. Consequently, the azimuthal intensity gradient increases since
\(
\frac{\partial I_{\pm}}
{\partial\phi}
\propto
|\ell_1-\ell_2|,
\)
leading to improved spatial resolution in the transverse plane. The radial index $m$ affects the radial intensity envelope but does not modify the
azimuthal factor responsible for the magnetic-field-dependent rotation. For a fixed
total probe power, increasing $m$ distributes the optical power among multiple radial
rings and therefore generally reduces the local intensity response at a fixed detection
point. Accordingly, we use $m=0$ in the analysis, while higher-$m$ modes
may be useful for spatial averaging when multiple rings are detected simultaneously.

Both detection schemes are subject to various noise sources. However, owing to its intrinsic phase stability, the common-path configuration suppresses most interferometric noise, making detector noise the dominant limiting factor. In contrast, the Mach--Zehnder configuration is highly sensitive to environmental perturbations, such as optical path length fluctuations and mechanical vibrations, which introduce additional interferometric noise. The influence of these noise sources on the magnetometer sensitivity is analyzed in the following section.


\begin{figure*}
\centering
\begin{subfigure}{.2\linewidth}
    \includegraphics[width=4cm]{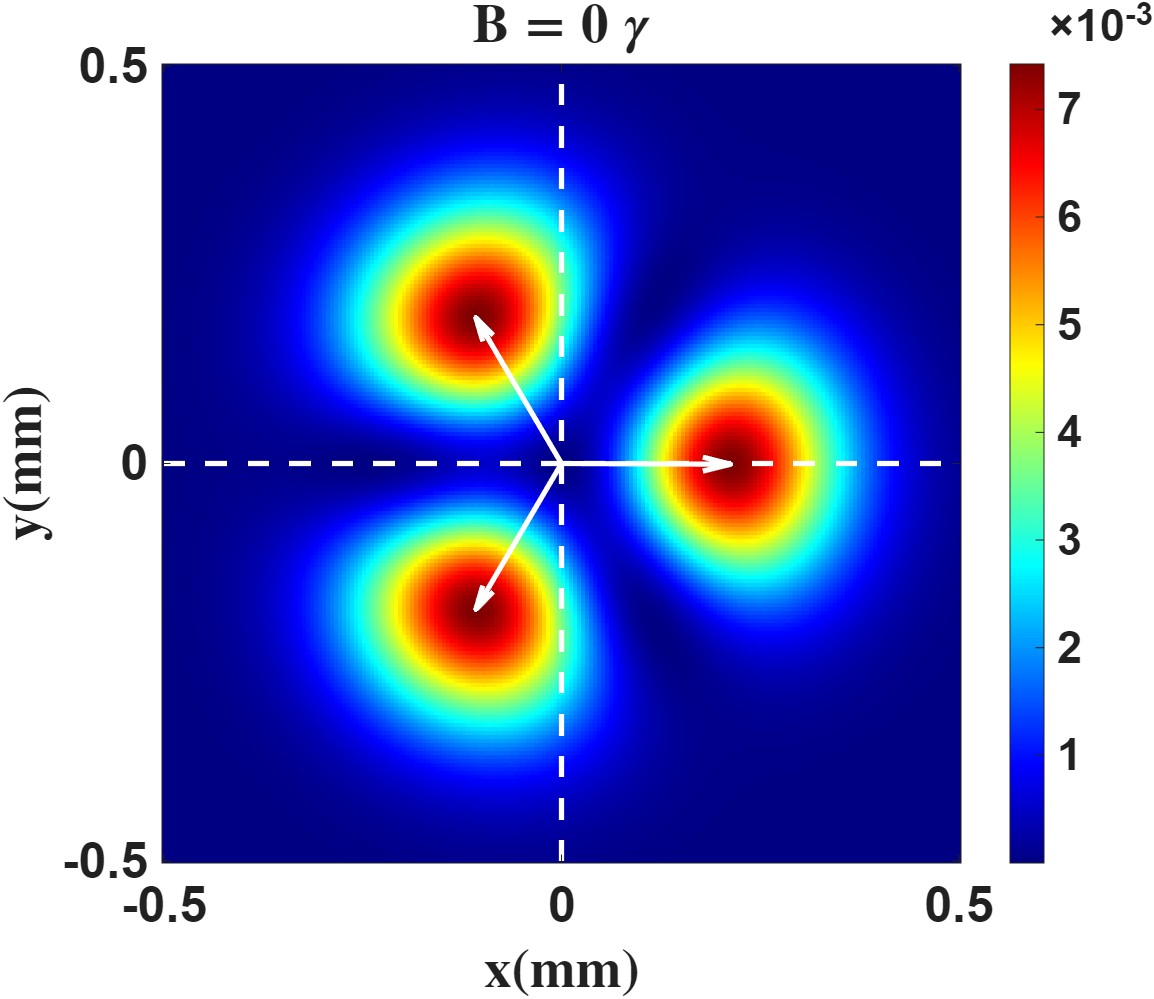}
    \caption{}
    \label{sp1}
\end{subfigure}
\hfill
\begin{subfigure}{.2\linewidth}
    \includegraphics[width= 4cm]{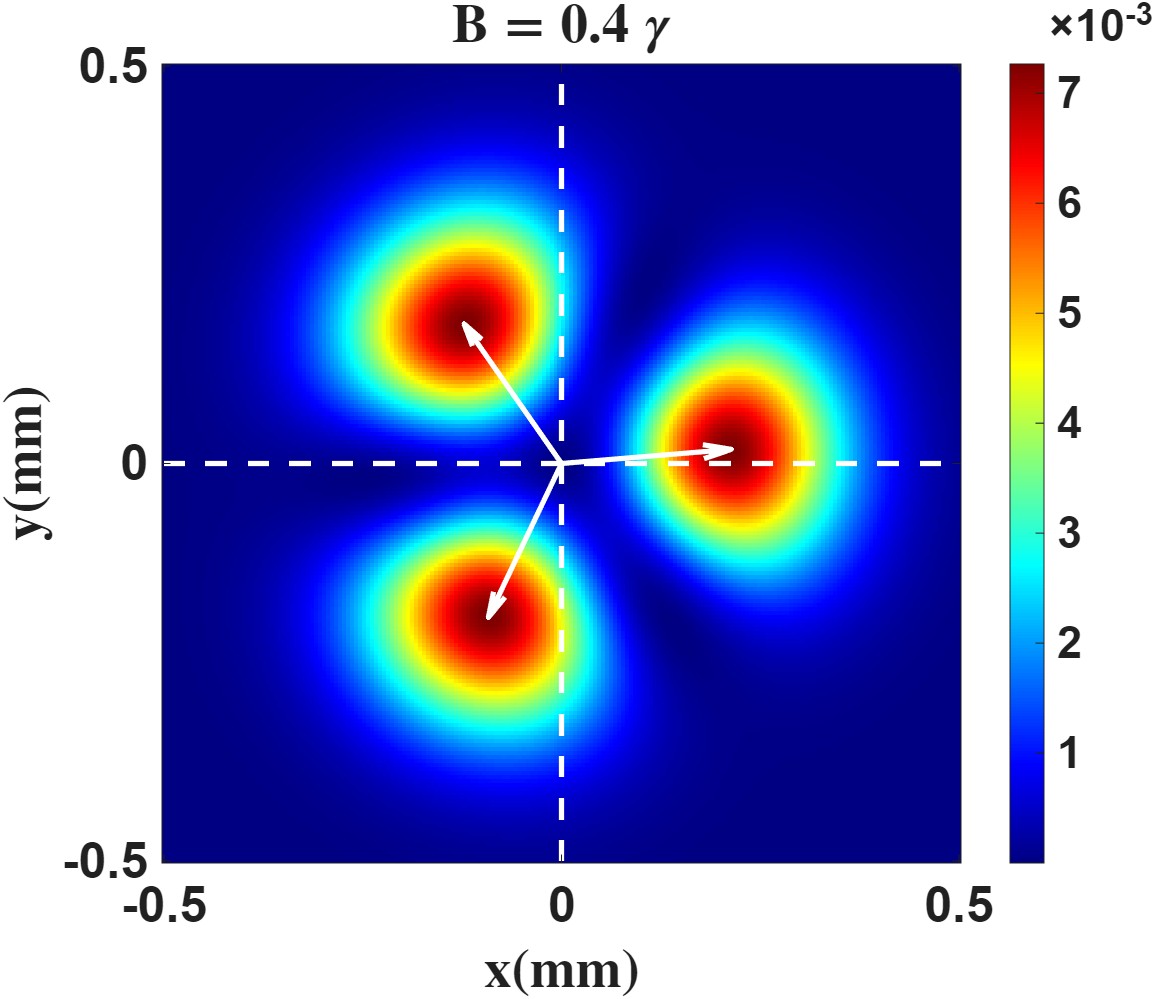}
    \caption{ }
    \label{sp2}
\end{subfigure}
\hfill
\begin{subfigure}{.2\linewidth}
    \includegraphics[width=4cm]{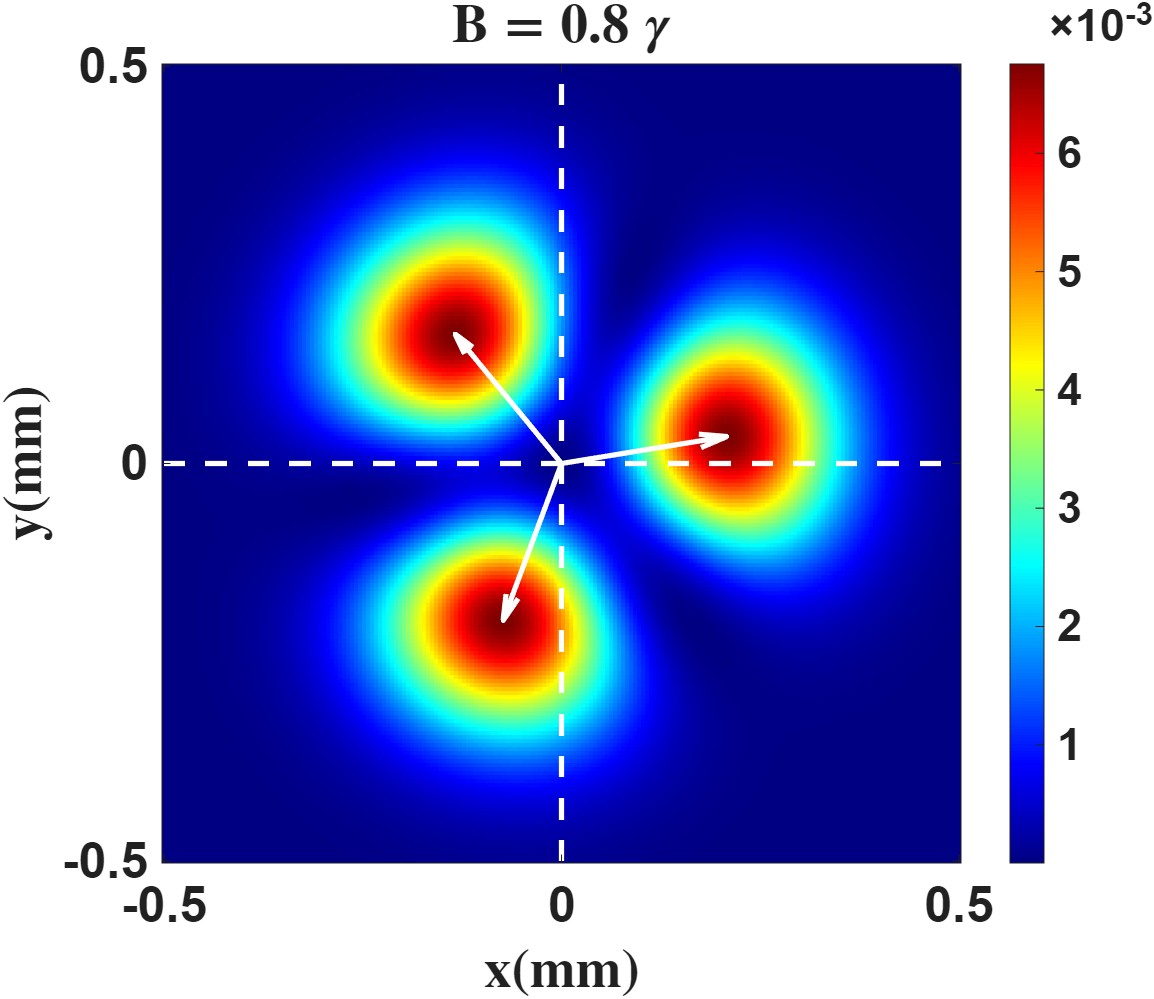}
    \caption{}
    \label{sp3}
\end{subfigure}
\hfill
\begin{subfigure}{.2\linewidth}
    \includegraphics[width= 4cm]{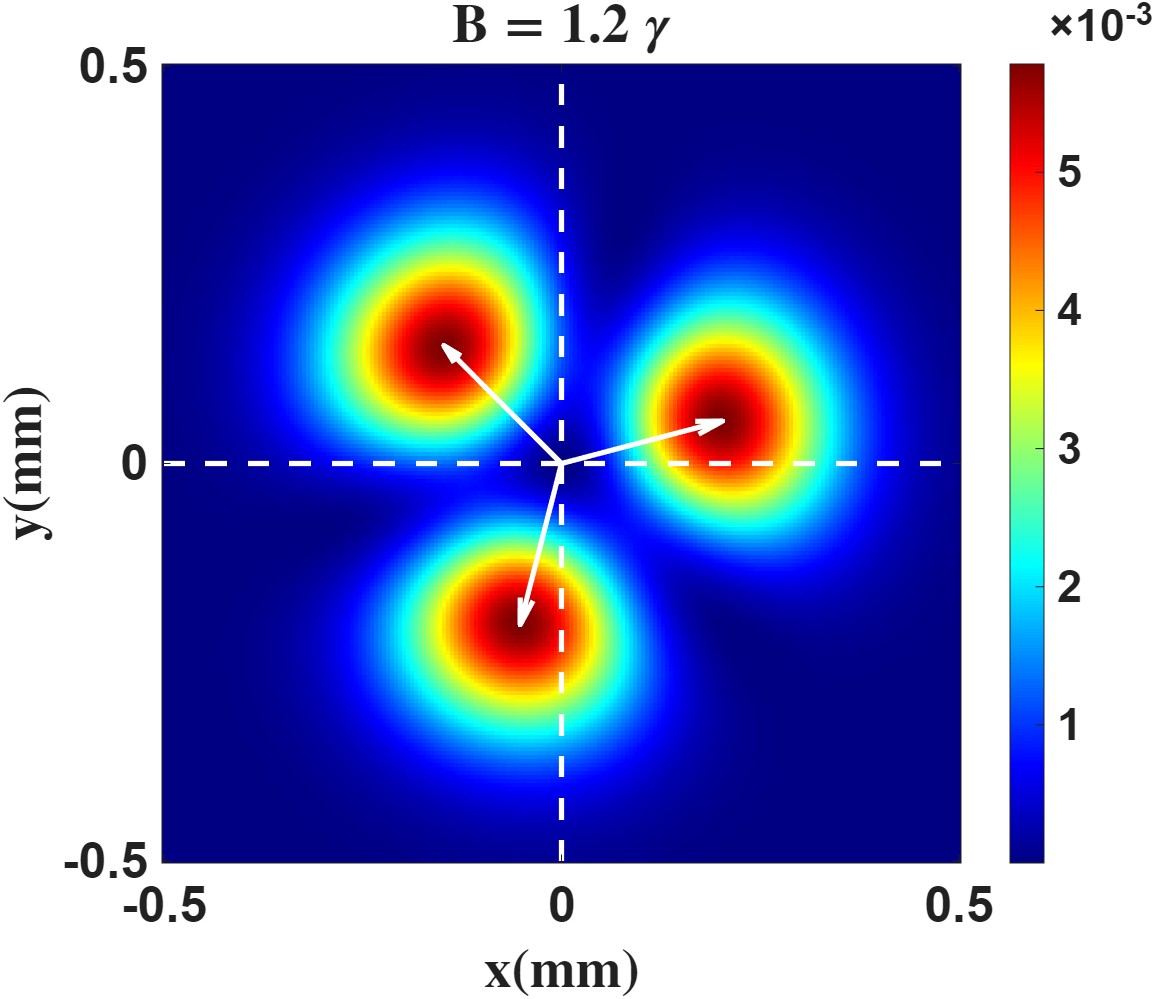}
    \caption{ }
    \label{sp4}
\end{subfigure}
\hfill
\begin{subfigure}{.2\linewidth}
    \includegraphics[width=4cm]{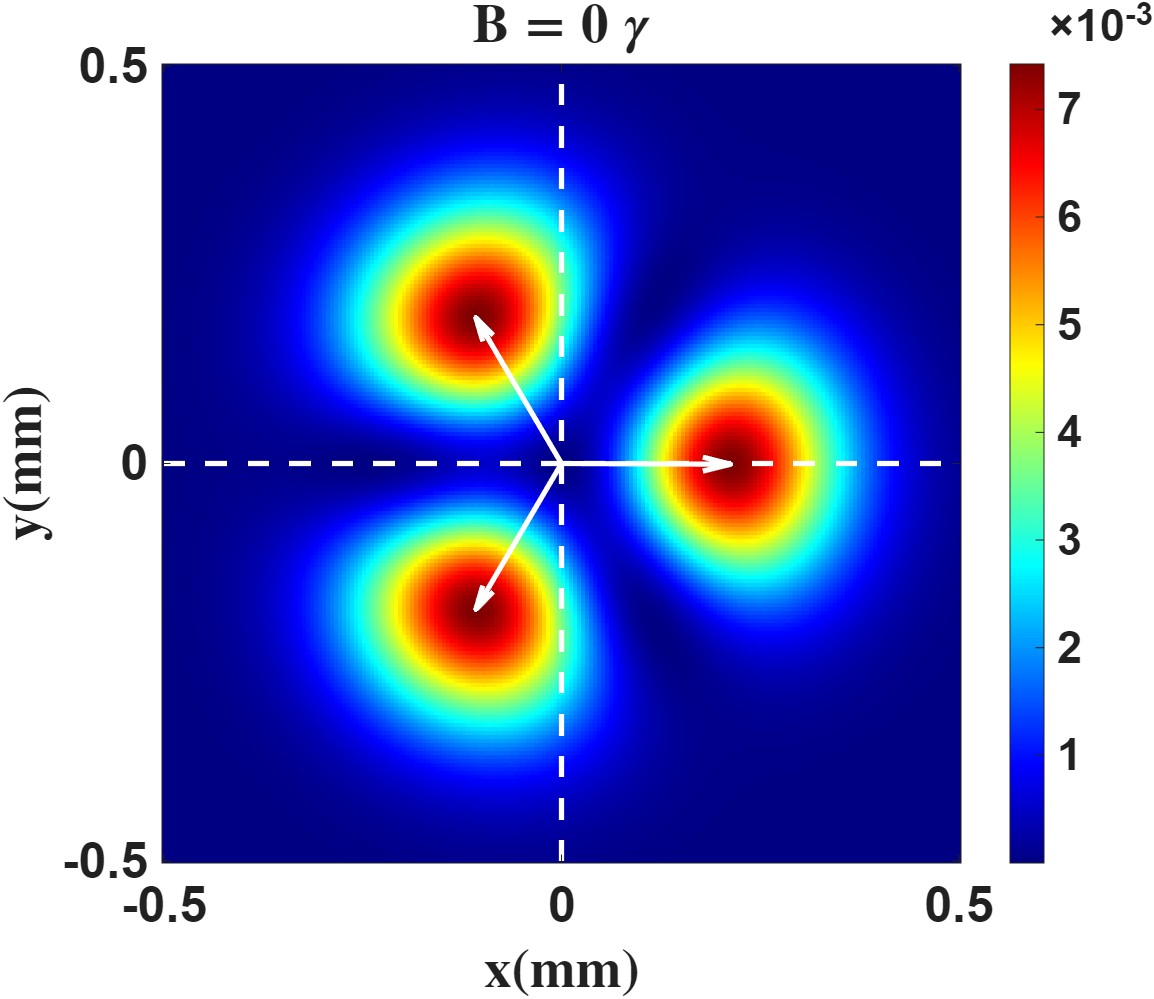}
    \caption{}
    \label{sp5}
\end{subfigure}
\hfill
\begin{subfigure}{.2\linewidth}
    \includegraphics[width= 4cm]{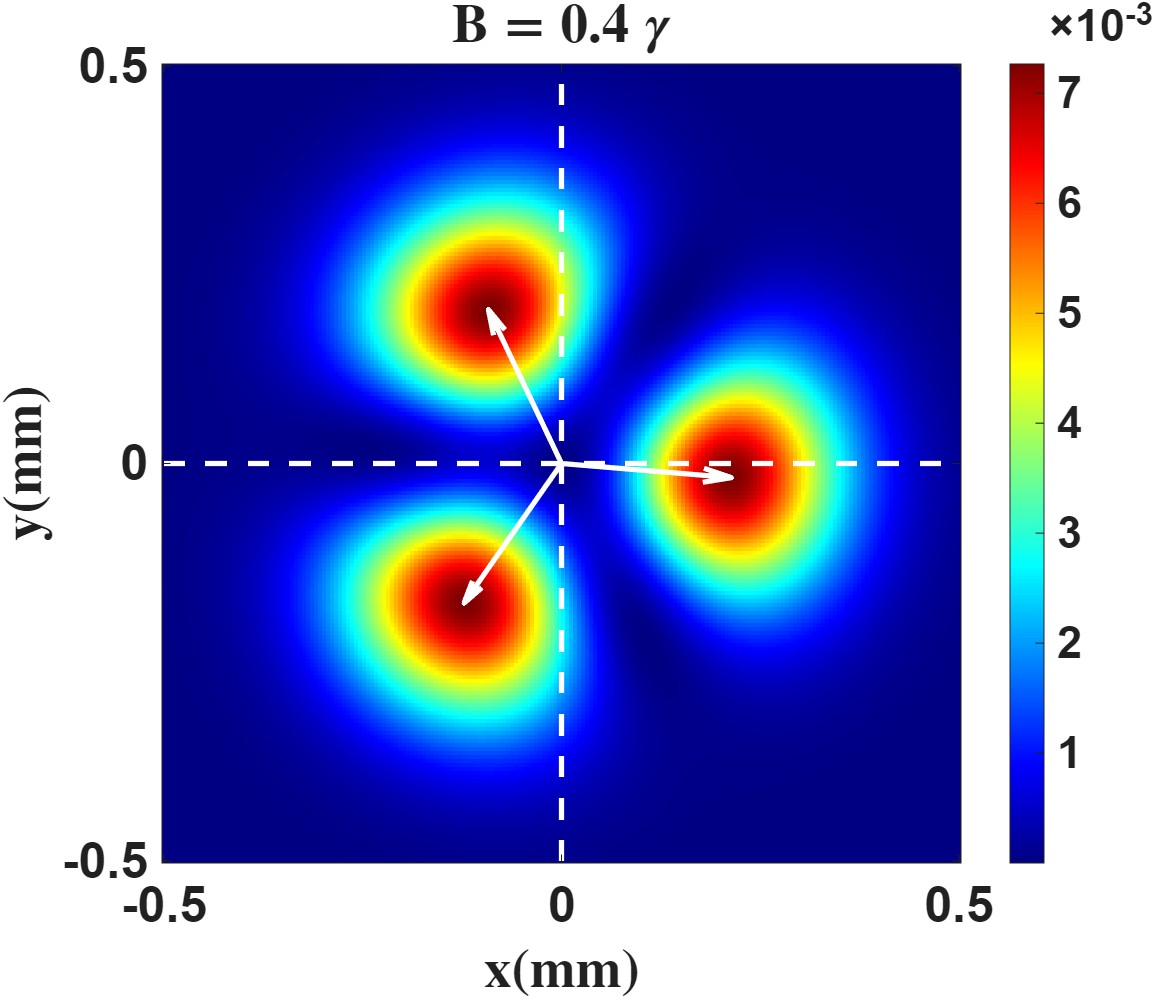}
    \caption{ }
    \label{sp6}
\end{subfigure}
\hfill
\begin{subfigure}{.2\linewidth}
    \includegraphics[width=4cm]{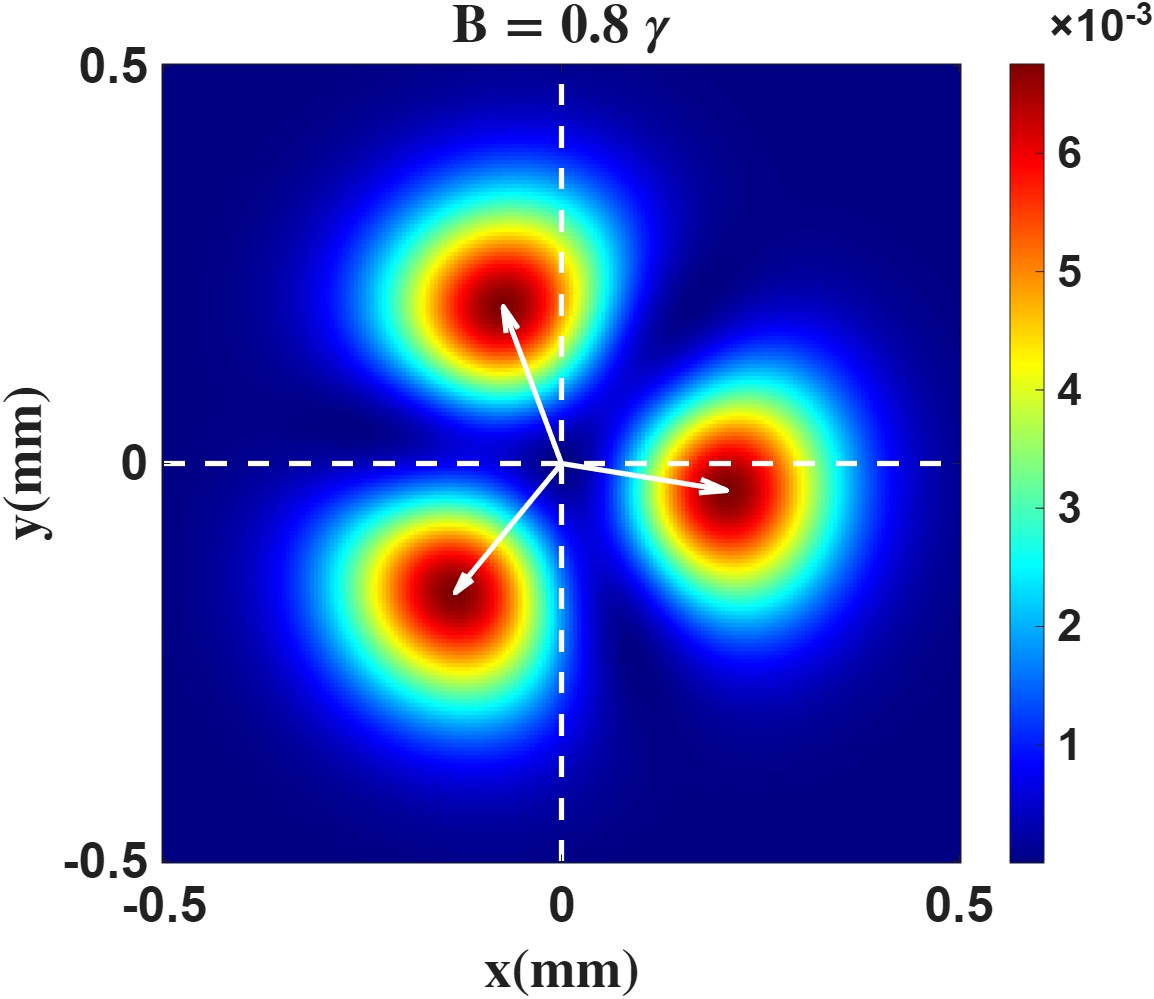}
    \caption{}
    \label{sp7}
\end{subfigure}
\hfill
\begin{subfigure}{.2\linewidth}
    \includegraphics[width= 4cm]{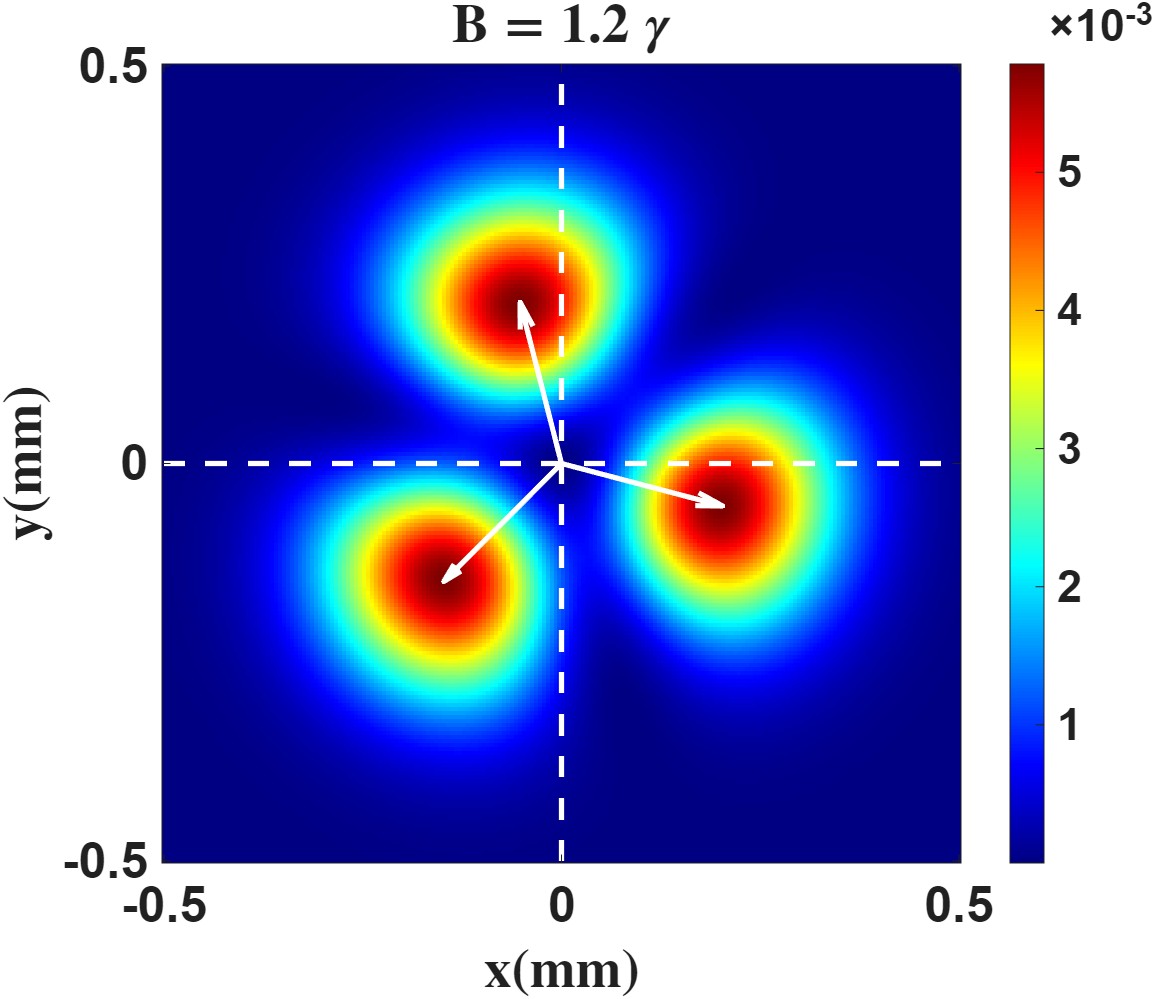}
    \caption{ }
    \label{sp8}
\end{subfigure}
\caption{Interference patterns \(I_+\) for (a)--(d) and the \(I_-\) for (e)--(h) at different magnetic-field strengths for $\ell_1=1$, $\ell_2=-2$. All other parameters are the same as those used in Fig.~\ref{com}.}
 \label{snp}
\end{figure*}

\begin{figure}
\centering
\begin{subfigure}{1\linewidth}
    \includegraphics[width=8cm]{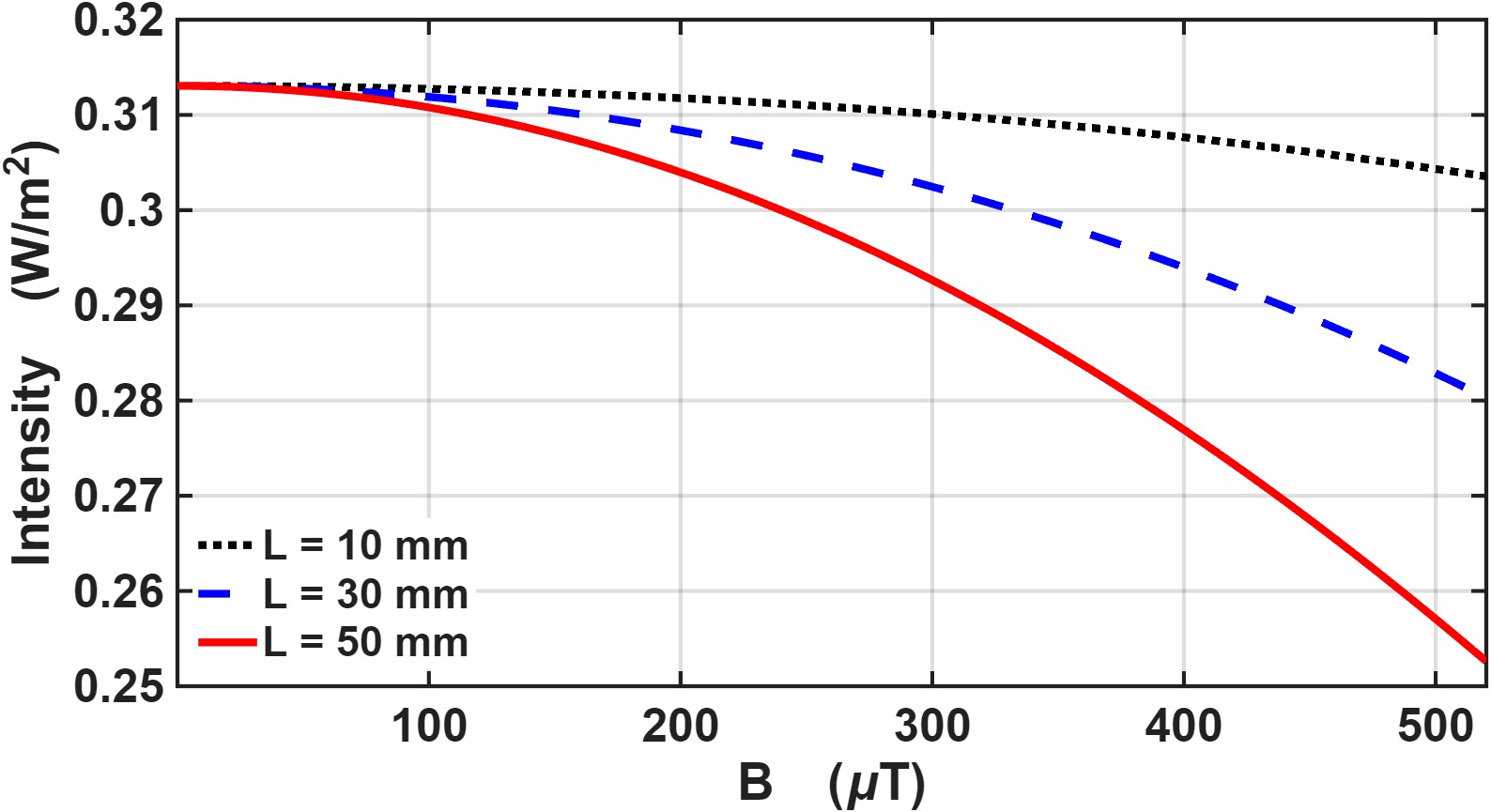}
    \caption{}
    \label{IvsB}
\end{subfigure}
\hfill
\begin{subfigure}{1\linewidth}
    \includegraphics[width= 8cm]{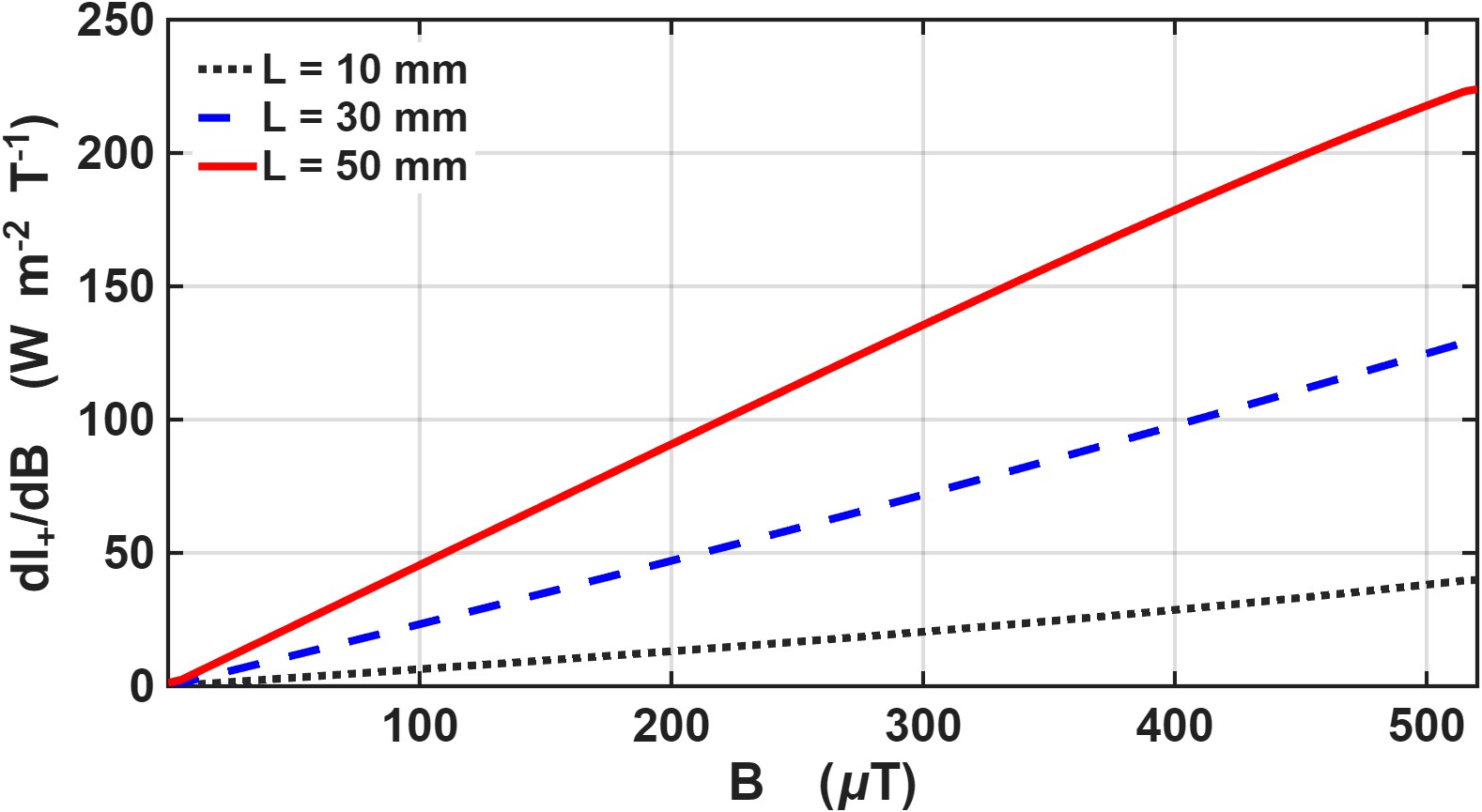}
    \caption{}
    \label{slopvar}
\end{subfigure}
\hfill
\caption{Intensity \(I_+\), evaluated at $x = w_0/\sqrt{2}$ and $y = 0$, plotted as functions of the magnetic field for length of the $L = 10$~mm (black dotted), $L = 30$~mm (blue dashed), and $L = 50$~mm (red solid), all other parameters are
those used in Fig. \ref{snp}.}
\label{IdI}
\end{figure}

\begin{figure*}
\centering
\begin{subfigure}{.2\linewidth}
    \includegraphics[width=4cm]{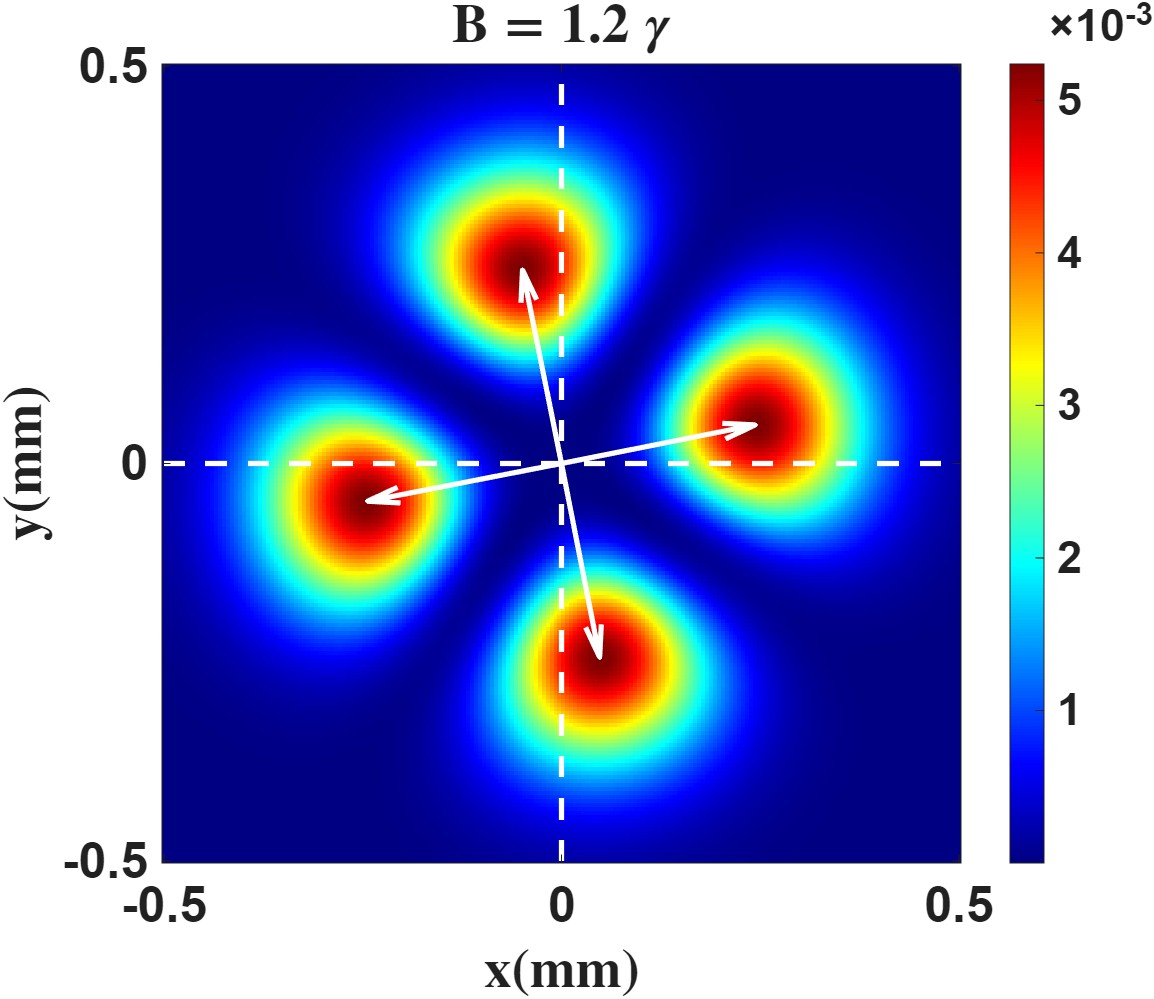}
    \caption{}
    \label{int4}
\end{subfigure}
\hfill
\begin{subfigure}{.2\linewidth}
    \includegraphics[width= 4cm]{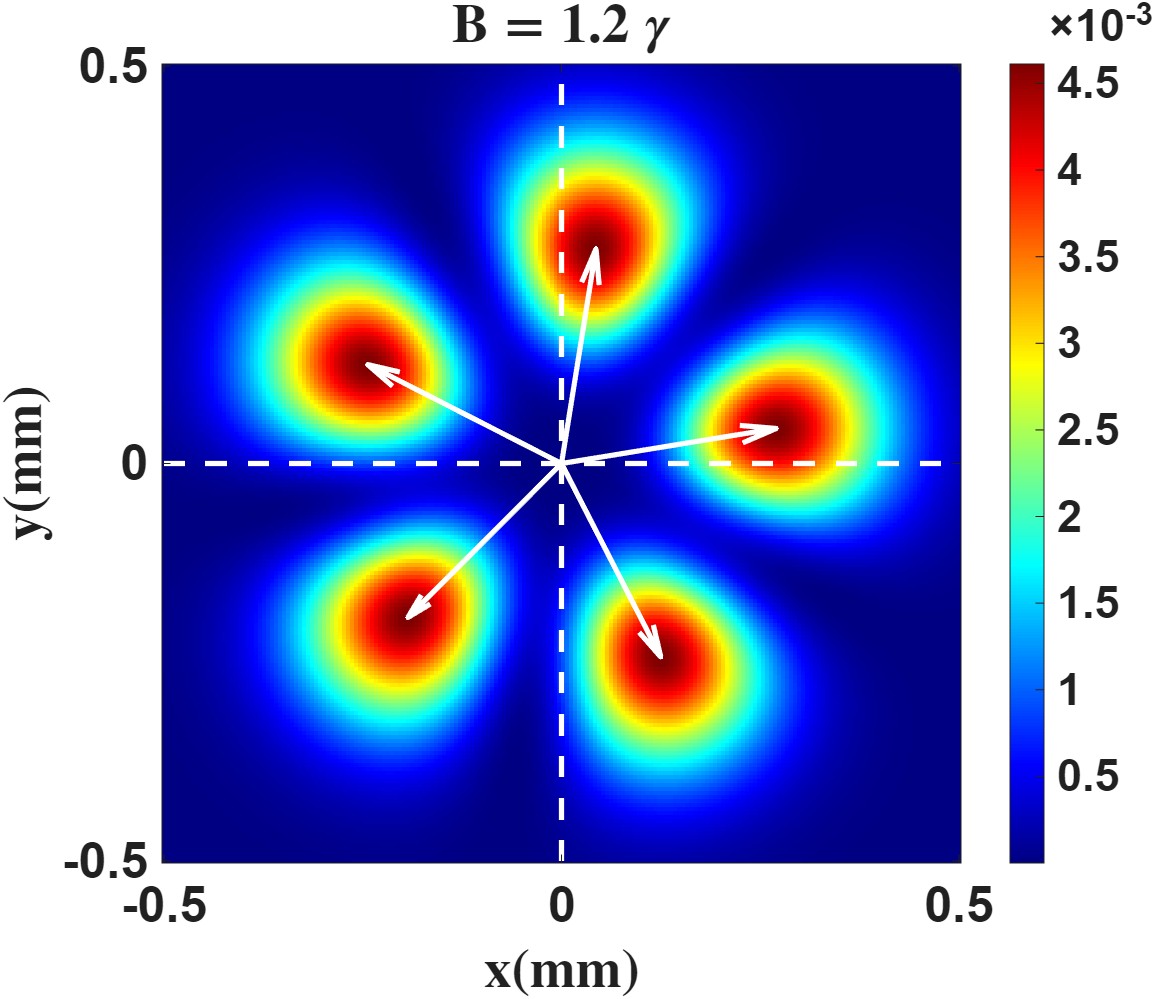}
    \caption{ }
    \label{int5}
\end{subfigure}
\hfill
\begin{subfigure}{.2\linewidth}
    \includegraphics[width=4cm]{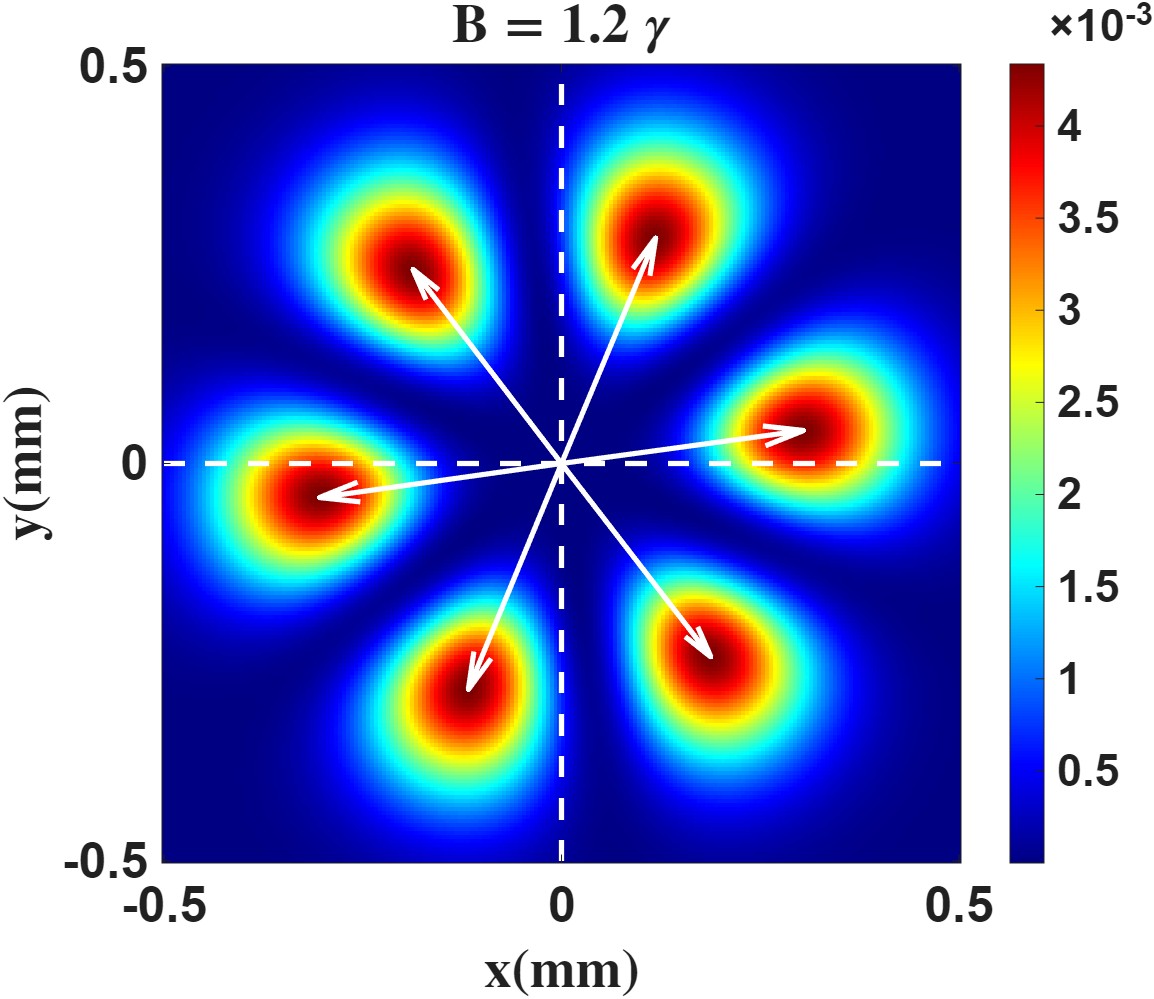}
    \caption{}
    \label{int6}
\end{subfigure}
\hfill
\begin{subfigure}{.2\linewidth}
    \includegraphics[width= 4cm]{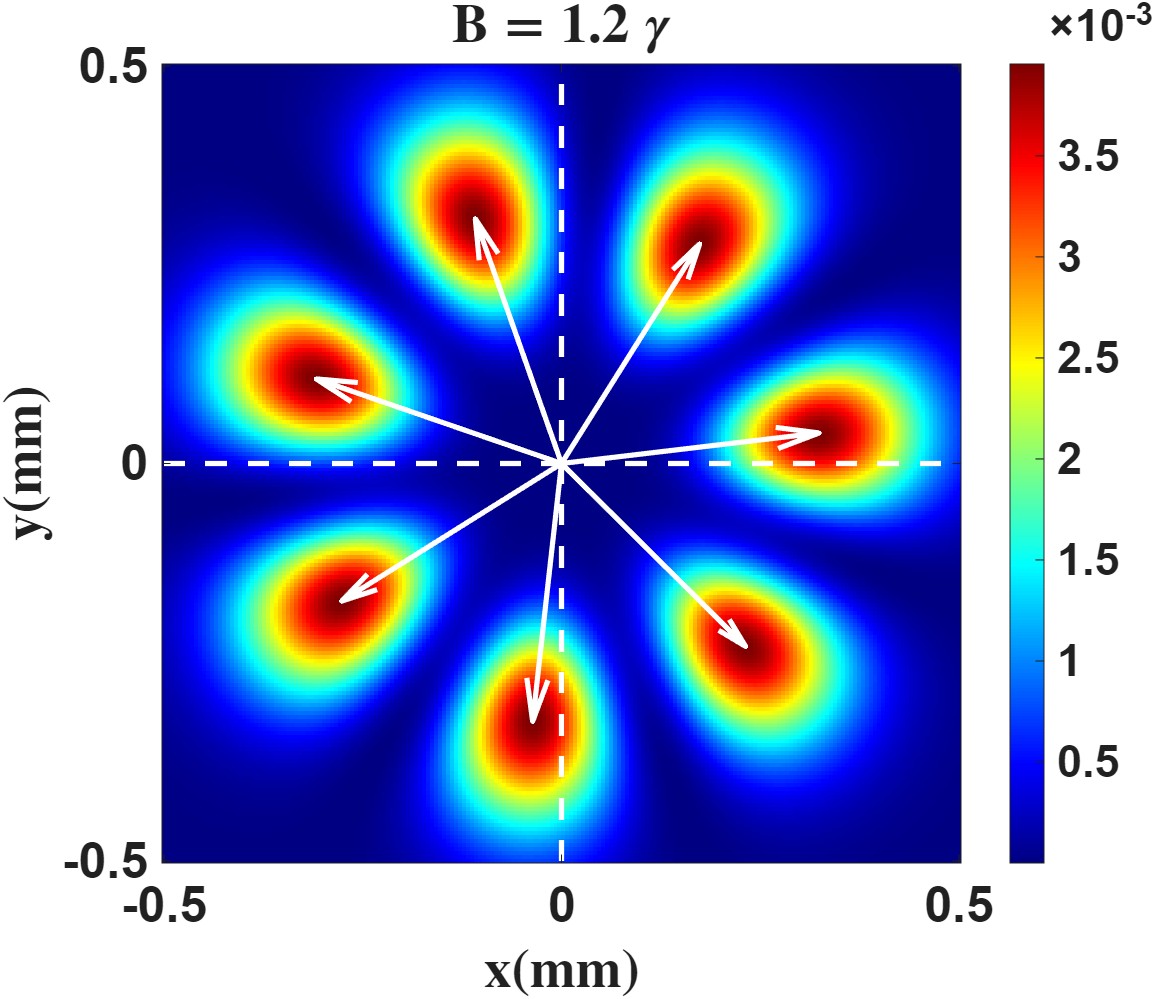}
    \caption{ }
    \label{int7}
\end{subfigure}
\label{int}
\caption{Interference patterns $I_{+}$ for different combinations of the OAM indices: (a) $\ell_{1}=2$, $\ell_{2}=-2$; (b) $\ell_{1}=3$, $\ell_{2}=-2$; (c) $\ell_{1}=3$, $\ell_{2}=-3$; and (d) $\ell_{1}=4$, $\ell_{2}=-3$. All other parameters are the same as those used in Fig.~\ref{com}.}
 \label{ints}
\end{figure*}

\section{Noise Analysis}

The ultimate sensitivity of the proposed interferometric magnetometer is limited by two independent noise sources: (i) interferometric phase noise arising from fluctuations in the optical path difference (OPD) between the signal and reference arms \cite{Musha:82}, and (ii) detector noise associated with the CCD camera used to record the interference pattern \cite{Irie_2008}. Assuming that these two noise sources are statistically independent, the total measurement uncertainty can be expressed as 
\begin{equation}
\delta I^{2}
=
\delta I_{\mathrm{int}}^{2}
+
\delta I_{\mathrm{cam}}^{2},
\label{eq:total_noise}
\end{equation}
where $\delta I_{\mathrm{int}}$ and $\delta I_{\mathrm{cam}}$ denote the standard deviations associated with interferometric and camera noise, respectively.

\subsubsection{Interferometric Phase Noise}

The interferometric detection scheme shown in Fig. \ref{mzi} is based on a MZI, in which the probe beam is divided into signal and reference beams by the first beam splitter ($BS_1$). The signal beam propagates through the optical vortex retarder ($OVR_1$), mirror ($M_1$), and the atomic medium before recombining with the reference beam at the second beam splitter ($BS_2$). The reference beam propagates through $OVR_2$ and mirror $M_2$ before interfering with the signal beam at $BS_2$. Since the two beams propagate along different optical paths, environmental perturbations acting differently on the two interferometer arms produce fluctuations in the measured phase difference.

The total phase accumulated by the signal arm is
\begin{equation}
\phi_s
=
\phi_{BS_1}^{s}
+
\phi_{OVR_1}
+
\phi_{M_1}
+
\phi_{air}^{s}
+
\phi_{atom}
+
\phi_{BS_2}^{s},
\label{eq:signal_phase}
\end{equation}
whereas the phase accumulated by the reference arm is
\begin{equation}
\phi_r
=
\phi_{BS_1}^{r}
+
\phi_{OVR_2}
+
\phi_{M_2}
+
\phi_{air}^{r}
+
\phi_{BS_2}^{r}.
\label{eq:reference_phase}
\end{equation}

Consequently, the phase difference recorded by the CCD camera is
\begin{equation}
\begin{aligned}
\Delta \Theta
&= \phi_s-\phi_r \\
&= \phi_{\mathrm{atom}}
+\delta\phi_{\mathrm{OVR}}
+\delta\phi_{\mathrm{BS}}
+\delta\phi_{\mathrm{M}}
+\delta\phi_{\mathrm{vib}}
+\delta\phi_{\mathrm{air}}, \end{aligned}
\label{eq:phase_difference}
\end{equation}
where \(\delta\phi_{OVR} =
(\ell_1-\ell_2)\phi\),
represents the azimuthal phase difference introduced by the two structured beams carrying OAM, and
\(\phi_{atom}
= \delta_{\pm}(B)\), denotes the magnetic-field-dependent phase shift induced by the atomic medium.

Since the beam splitters \(BS_1\) and \(BS_2\), as well as the mirrors \(M_1\) and \(M_2\), are assumed to be identical, they do not introduce any systematic differential phase shift, i.e.,
\(\delta\phi_{BS}=0\) and \(\delta\phi_{M}=0\). However, mechanical vibrations of the mirrors, beam splitters, optical mounts, and other interferometric components can induce fluctuations in the optical path length between the two arms of the interferometer. The corresponding phase fluctuation can be expressed in terms of the differential optical path-length variation as
\begin{equation}
\delta\phi_{\mathrm{vib}}
=
\frac{2\pi n_{air}}{\lambda}\,
\delta L_{\mathrm{vib}},
\label{eq:vibration_phase}
\end{equation}
where \(n_{air}\) is the refractive index of the air, \(\lambda\) is the wavelength of the probe beam, and \(\delta L_{\mathrm{vib}}\) represents the vibration-induced differential optical path-length fluctuation between the two interferometer arms. Assuming that the optical path length fluctuation due to mechanical vibrations, $\delta L_{\mathrm{vib}}$, is of the order of nanometres for unit bandwidth \cite{NBobroff_1993}, the corresponding phase error is approximately 8.05$\times10^{-3}\,\mathrm{rad}/\mathrm{\sqrt{Hz}}$ . Since this phase error contributes significantly to the overall measurement uncertainty, the influence of mechanical vibrations must be minimized through appropriate vibration isolation and stabilization techniques.

Temperature fluctuations modify the refractive index of air, thereby introducing unequal optical phases in the two interferometer arms. The corresponding phase fluctuation can be written as
\begin{equation}
\delta\phi_{\mathrm{air}}
=
\frac{2\pi}{\lambda}
(L_s -
L_r)
\frac{dn_{air}}{dT}
\delta T.
\label{eq:air_phase}
\end{equation}

Here, \(L_s\) and \(L_r\) represent the free-space propagation lengths of the signal and reference arms, respectively, while \(dn_{\mathrm{air}}/dT\) denotes the thermo-optic coefficient of air, whose magnitude is typically of the order of \(10^{-6}\,\mathrm{K}^{-1}\) \cite{NBobroff_1993}. The quantity \(\delta T\) accounts for the temperature fluctuations experienced by the two optical paths. Assuming a temperature drift of the order of \(10^{-4}\,\mathrm{K}\) \cite{10.1117/1.OE.54.10.104104} for unit bandwidth, which is typical of a standard air-conditioned laboratory, and considering a path-length difference of \(L_s-L_r=50~\mathrm{mm}\), the resulting phase fluctuation due to the temperature-dependent variation of the refractive index of air is given by
\(\delta\phi_{\mathrm{air}} \approx 4.03\times10^{-5}\)rad/\(\sqrt{\rm Hz}\). The estimated value of \(\delta\phi_{\mathrm{air}}\) is sufficiently small compared with the phase shift induced by the magnetic field, indicating that the influence of air-induced phase fluctuations is negligible under typical laboratory conditions. 

The interferometric intensity fluctuation was obtained by evaluating the difference between the interference intensity in the presence and absence of the phase fluctuation at fixed, namely,
\begin{equation}
\delta I_{\rm int}
=
\left|I(\Theta+\delta\Theta)-I(\Theta)\right|,
\end{equation}
where \(\delta\Theta = \delta \phi_{vib}+\delta \phi_{air}\). In the limit \(\delta\Theta\ll1\), the above expression reduces to \(\delta I_{\rm int}\simeq I_0 |\sin\Theta|\,\delta\Theta.\) the corresponding interference intensity fluctuation at $B = 80\,\mu\mathrm{T}$ is $\delta I_{\mathrm{int}} = 4.2 \times 10^{-4}\,\mathrm{W\,m^{-2}}/\sqrt{\rm Hz}$.

In the present work, fluctuations associated with the atomic medium, including variations in atomic density, laser intensity, and laser frequency, are neglected. This approximation is justified because the MOR angle is determined by the differential susceptibility of the two co-propagating circularly polarized components, namely,
\(\theta \propto \mathrm{Re}\left(\chi_{-}-\chi_{+}\right).\)
Consequently, fluctuations that affect both polarization components in an identical manner appear primarily as common-mode noise and are therefore strongly suppressed in the differential measurement process. Residual differential contributions arising from laser-frequency fluctuations, laser-intensity fluctuations, and atomic-density variations are assumed to be sufficiently small and are neglected in the present analysis \cite{Li:08}.

\subsubsection{Camera Noise}
A CCD camera records the interference pattern emerging from \(BS_2\). If \(N_{\mathrm{ph}}\) photons per second are incident on the detector during the exposure time, the corresponding number of detected photoelectrons is given by
\begin{equation}
N_e=\eta N_{\mathrm{ph}},
\end{equation}
where \(\eta\) is the quantum efficiency of the CCD camera \cite{CCD_SNR, Sperlich_2014}.

The detector noise consists of three statistically independent contributions: (i) photon shot noise, (ii) readout noise, and (iii) dark-current noise. Consequently, the total root-mean-square (RMS) uncertainty in the detected photoelectrons is
\begin{equation}
\delta N=\sqrt{N_e+R^2+Dt},
\end{equation}
where \(R\) denotes the readout noise (electrons RMS/pixel), \(D\) is the dark-current generation rate (electrons/pixel/s), and \(t\) is the camera exposure time. The first term represents the Poissonian photon shot noise, whose variance is equal to the mean number of detected photoelectrons.

For the probe field strength \(\Omega_p=0.2\gamma\), the total number of incident photons is approximately \(N_{\mathrm{ph}}\)=6.4\(\times\)\(10^{10}\) per second. Considering a beam waist of \(w_0=250~\mu\mathrm{m}\) and a CCD pixel size of \(15~\mu\mathrm{m}\), the beam illuminates approximately \(N_{\mathrm{pix}}\approx 1 \times 10^3\) pixels. Assuming a nearly uniform photon distribution over the illuminated area, the average number of detected photoelectrons per pixel is therefore
\begin{equation}
N_{e,\mathrm{pix}}
=\frac{\eta N_{\mathrm{ph}}}{N_{\mathrm{pix}}}.
\end{equation}

For a detector quantum efficiency of \(\eta=0.8\), the detected photoelectron rate per pixel is estimated to be \(N_{e,\mathrm{pix}}=5\times10^{7}~\mathrm{s}^{-1}\). Since the atomic system reaches a steady state within a few microseconds to milliseconds \cite{Steck2021}, the required exposure time is also of the same order. This timescale is considerably shorter than the typical pixel saturation time \cite{CCD_SNR}, indicating that detector saturation can be neglected.
Since
\(
N_{e,\mathrm{pix}} \gg R\), \(N_{e,\mathrm{pix}} \gg D,
\)
where $R \approx 2\!-\!5$ electrons RMS per pixel and
$D \approx 10\!-\!20$ electrons per pixel per second for a typical scientific CCD camera, the detector noise contributions are negligible.  Therefore, the detector operates in the photon shot-noise-limited regime \cite{budker2013optical}, and the detector noise can be approximated \(\delta N \approx \sqrt{N_{e,\mathrm{pix}}}.\)
The corresponding camera-induced intensity noise is estimated to be \(\delta I_{\mathrm{cam}} = 5.7\times10^{-6}~\mathrm{W\,m^{-2}\,pixel^{-1}\,/\sqrt{Hz}}\).

Under the noise limit, the magnetic-field sensitivity $\delta B$ is given by \cite{budker2013optical}
\begin{equation}
\delta B= \delta I
\left(\frac{\partial I}{\partial B}\right)^{-1}.
\label{eq:sensitivity}
\end{equation}
 Since the analysis is performed in the EIT regime, absorption is strongly suppressed, and the detected photon number remains nearly equal to the incident photon flux. Consequently, improving sensitivity primarily requires maximizing the differential angular response $\partial I/\partial B$.

The rotation of the interference pattern originates from magnetically induced birefringence. The applied magnetic field produces Zeeman splitting, while the control field $\Omega_c$ modifies the dispersive properties of the medium through dressed-state formation. As shown in Fig.~\ref{rho}, the dispersive response exhibits a linear region whose width and slope depend sensitively on $\Omega_c$. For small $\Omega_c$, the differential dispersion varies rapidly with magnetic field, producing a large rotation slope, whereas increasing $\Omega_c$ broadens the transparency window and reduces the dispersion gradient.
A critical regime arises when the Zeeman splitting becomes comparable to the dressed-state splitting ($B \sim \Omega_c$). In this regime, absorption increases, and the refractive index decreases, making the reliable determination of the MOR angle difficult. Therefore, optimal operation must remain within the dispersion-dominated EIT region.

In the previous section, the MOR angle was derived from the interference intensity of a radially polarized probe beam [Eq.~(\ref{16})]. Since the output intensity depends on the atomic coherences, the rotation angle varies with $\Omega_c$, $B$, and length of the medium $L$. Contour plot of \(\partial I_+/\partial B\) [Fig.~\ref{dIdBcont}] at $L=50$ mm show that rotation is maximized for moderate control-field strengths. 
 Smaller values of $\Omega_c$  produce steeper dispersion and higher sensitivity, whereas larger values of $\Omega_c$ broaden the response region but reduce the slope. Extremely small $\Omega_c$, however, degrade EIT and reduce interference visibility, implying the existence of an optimal control-field strength.


For an interaction length of \(L=50~\mathrm{mm}\) and a probe-field strength of \(0.2\gamma\),  a moderate strength of the control field \(\Omega_c\) yields the value of \(\partial I_{+}/\partial B \approx 6\times10^{2}~\mathrm{W\,m^{-2}}\) [using Fig.~\ref{dIdBcont}]. Accounting for both the interferometric fluctuations, \(\delta I_{\mathrm{int}}\), and the camera noise, \(\delta I_{\mathrm{cam}}\), the resulting magnetic-field sensitivity is estimated to be \(7.0\times10^{-7}~\mathrm{T}/\sqrt{\mathrm{Hz}}\). This demonstrates the importance of optimizing the control-field strength \(\Omega_c\) to maximize the sensitivity of the measurement scheme. In contrast, for a common-path interferometric arrangement, the effect of interferometric phase noise is substantially reduced, and the sensitivity is primarily limited by the camera noise, yielding an estimated sensitivity of \(9.5~\mathrm{nT}/\sqrt{\mathrm{Hz}}\).

Although the MZI configuration requires effective suppression of mechanical vibrations and environmental phase noise to achieve high magnetic-field sensitivity, it offers a significant advantage in spatially resolved magnetometry. In particular, it enables imaging of spatially varying magnetic fields, where increasing the topological charge difference, $|\ell_1-\ell_2|$, enhances the azimuthal intensity gradient, $\partial I_{\pm}/\partial \phi \propto |\ell_1-\ell_2|$, thereby improving the spatial encoding of polarization rotation. Furthermore, the control-field strength serves as a key parameter for tailoring the interference intensity and, consequently, the magnetic-field response. In the present work, the control field is assumed to be spatially uniform; however, employing a structured control field could provide an additional degree of freedom for engineering the spatial coherence and optimizing the sensitivity. This represents a promising direction for future research. The key novelty of the proposed approach lies in the spatial transduction of magnetically induced polarization rotation into a directly observable interference pattern, enabling visually intuitive and spatially resolved magnetometry at finite magnetic fields.
\begin{figure}
\centering
    \includegraphics[width=8cm]{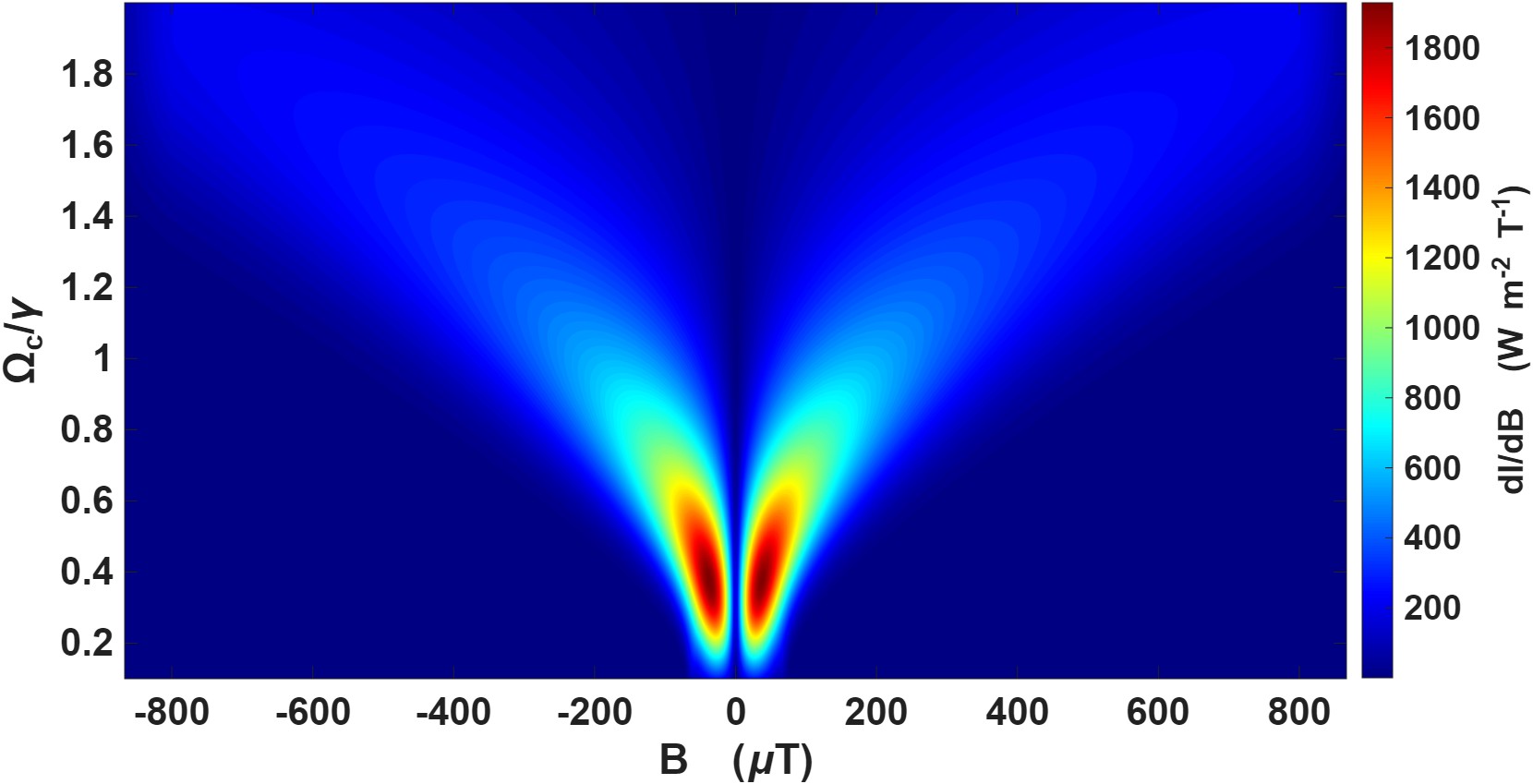}
   
\caption{Contour plot of the slope of the intensity  \(I_+\) with respect to $B$, $\partial I_+ / \partial B$, evaluated at propagation distances $z = 50$ $\mathrm{mm}$, as functions of the normalized control field strength $\Omega_c/\gamma$ and magnetic field $B/\gamma$. All other parameters are the same as those used in Fig.\ref{com}.}
 \label{dIdBcont}
\end{figure}

\section{Conclusion}
In this work, we investigate the interaction of a radially polarized LG beam with an atomic medium subjected to a static magnetic field. The magnetic field induces differential dispersion between the $\sigma_+$ and $\sigma_-$ components, resulting in MOR. Since the accumulated phase is not directly observable in the beam intensity, we employ a common-path interferometric configuration as the primary detection scheme, where the $\sigma_+$ and $\sigma_-$ components are coherently superposed. We further extend this approach to an MZI using a reference radially polarized LG beam, providing independent control over the OAM difference and hence the number of interference petals.

Our results demonstrate that structured light provides a direct link between OAM and the spatial resolution of magnetic-field measurements. Increasing the OAM difference increases the number of interference petals and enhances the transverse spatial structure, making the scheme particularly suitable for imaging spatially varying magnetic fields. The sensitivity can be further optimized through the control-field strength and interaction length of the atomic medium.

The proposed approach converts magnetically induced polarization rotation into a measurable spatial displacement of the interference pattern, avoiding balanced photodetection and full Stokes-vector reconstruction. The common-path configuration provides a stable and sensitive primary detection method, while the MZI extension offers additional control over the spatial structure and resolution. These features highlight the potential of structured-light interferometry for precision magnetometry and spatially resolved magnetic-field sensing.

\begin{acknowledgments}
One of us (P.B.) acknowledges the financial support provided by the Council for Scientific and Industrial Research (CSIR), India, during this work.
\end{acknowledgments}

\bibliography{apssamp}

@PREAMBLE{
 "\providecommand{\noopsort}[1]{}" 
 # "\providecommand{\singleletter}[1]{#1}%" 
}

@article{TAMBAG2023129649,
title = {Direct observation of the Faraday rotation using radially-polarised doughnut modes},
journal = {Optics Communications},
volume = {545},
pages = {129649},
year = {2023},
issn = {0030-4018},
doi = {https://doi.org/10.1016/j.optcom.2023.129649},
url = {https://www.sciencedirect.com/science/article/pii/S0030401823003978},
author = {F. Tambag and K. Koksal and F. Yildiz and M. Babiker}
}

@article{PhysRevA.93.063826,
  title = {Probing vacuum-induced coherence via magneto-optical rotation in molecular systems},
  author = {Kumar, Pardeep and Deb, Bimalendu and Dasgupta, Shubhrangshu},
  journal = {Phys. Rev. A},
  volume = {93},
  issue = {6},
  pages = {063826},
  numpages = {11},
  year = {2016},
  month = {Jun},
  publisher = {American Physical Society},
  doi = {10.1103/PhysRevA.93.063826},
  url = {https://link.aps.org/doi/10.1103/PhysRevA.93.063826}
}

@ARTICLE{1,
  
AUTHOR={Bai, Xuanyao  and Wen, Kailun  and Peng, Donghong  and Liu, Shuangqiang  and Luo, Le },
         
TITLE={Atomic magnetometers and their application in industry},
        
JOURNAL={Frontiers in Physics},
        
VOLUME={Volume 11 - 2023},

YEAR={2023},

URL={https://www.frontiersin.org/journals/physics/articles/10.3389/fphy.2023.1212368},

DOI={10.3389/fphy.2023.1212368},

ISSN={2296-424X}}

@article{AN2022103752,
title = {Imaging somatosensory cortex responses measured by OPM-MEG: Variational free energy-based spatial smoothing estimation approach},
journal = {iScience},
volume = {25},
number = {2},
pages = {103752},
year = {2022},
issn = {2589-0042},
doi = {https://doi.org/10.1016/j.isci.2022.103752},
url = {https://www.sciencedirect.com/science/article/pii/S2589004222000220},
author = {Nan An and Fuzhi Cao and Wen Li and Wenli Wang and Weinan Xu and Chunhui Wang and Min Xiang and Yang Gao and Binbin Sui and Aimin Liang and Xiaolin Ning}
}

@article{10.1116/5.0025186,
    author   = {Fu, Kai-Mei C. and Iwata, Geoffrey Z. and Wickenbrock, Arne and Budker, Dmitry},
    title    = {Sensitive magnetometry in challenging environments},
    journal  = {AVS Quantum Science},
    volume   = {2},
    number   = {4},
    pages    = {044702},
    year     = {2020},
    month    = {12},
    issn     = {2639-0213},
    doi      = {10.1116/5.0025186},
    url      = {https://doi.org/10.1116/5.0025186}
}

@article{Bennett2021-im,
  title    = {Precision Magnetometers for Aerospace Applications: A Review},
  author   = {Bennett, James S and Vyhnalek, Brian E and Greenall, Hamish and
              Bridge, Elizabeth M and Gotardo, Fernando and Forstner, Stefan
              and Harris, Glen I and Miranda, F{\'e}lix A and Bowen, Warwick P},
  journal  = {Sensors (Basel)},
  volume   =  21,
  number   =  16,
  month    =  aug,
  year     =  2021,
  address  = {Switzerland},
}

@Article{Aslam2023,
author={Aslam, Nabeel
and Zhou, Hengyun
and Urbach, Elana K.
and Turner, Matthew J.
and Walsworth, Ronald L.
and Lukin, Mikhail D.
and Park, Hongkun},
title={Quantum sensors for biomedical applications},
journal={Nature Reviews Physics},
year={2023},
month={Mar},
day={01},
volume={5},
number={3},
pages={157-169},
issn={2522-5820},
doi={10.1038/s42254-023-00558-3},
url={https://doi.org/10.1038/s42254-023-00558-3}
}

@ARTICLE{10.3389/fphy.2023.1212368,
  
AUTHOR={Bai, Xuanyao  and Wen, Kailun  and Peng, Donghong  and Liu, Shuangqiang  and Luo, Le },
         
TITLE={Atomic magnetometers and their application in industry},
        
JOURNAL={Frontiers in Physics},
        
VOLUME={Volume 11 - 2023},

YEAR={2023},

URL={https://www.frontiersin.org/journals/physics/articles/10.3389/fphy.2023.1212368},

DOI={},

ISSN={2296-424X}}

@article{RevModPhys.92.015004,
  title = {Sensitivity optimization for NV-diamond magnetometry},
  author = {Barry, John F. and Schloss, Jennifer M. and Bauch, Erik and Turner, Matthew J. and Hart, Connor A. and Pham, Linh M. and Walsworth, Ronald L.},
  journal = {Rev. Mod. Phys.},
  volume = {92},
  issue = {1},
  pages = {015004},
  numpages = {68},
  year = {2020},
  month = {Mar},
  publisher = {American Physical Society},
  doi = {10.1103/RevModPhys.92.015004},
  url = {https://link.aps.org/doi/10.1103/RevModPhys.92.015004}
}

@ARTICLE{10963667,
  author={Marsic, Vlad and Igic, Petar and Pandey, Vartika and Zhou, Wenzhi and Faramehr, Soroush},
  journal={IEEE Access}, 
  title={Experimental Low-Cost Magnetic Scanning for Non-Destructive Testing With Hall Sensors: Determining Material’s Thickness}, 
  year={2025},
  volume={13},
  number={},
  pages={65105-65120},
  doi={10.1109/ACCESS.2025.3560025}}

@article{10.1063/1.5001730,
    author   = {Sheng, Jingwei and Wan, Shuangai and Sun, Yifan and Dou, Rongshe and Guo, Yuhao and Wei, Kequan and He, Kaiyan and Qin, Jie and Gao, Jia-Hong},
    title    = {Magnetoencephalography with a Cs-based high-sensitivity compact atomic magnetometer},
    journal  = {Review of Scientific Instruments},
    volume   = {88},
    number   = {9},
    pages    = {094304},
    year     = {2017},
    month    = {09},
    issn     = {0034-6748},
    doi      = {10.1063/1.5001730},
    url      = {https://doi.org/10.1063/1.5001730}
}

@article{10.1063/1.2354545,
    author = {Fagaly, R. L.},
    title = {Superconducting quantum interference device instruments and applications},
    journal = {Review of Scientific Instruments},
    volume = {77},
    number = {10},
    pages = {101101},
    year = {2006},
    month = {10},
    issn = {0034-6748},
    doi = {10.1063/1.2354545},
    url = {https://doi.org/10.1063/1.2354545},
    }

@ARTICLE{Wei2021-cl,
  title    = "Recent Progress of Fluxgate Magnetic Sensors: Basic Research and
              Application",
  author   = "Wei, Songrui and Liao, Xiaoqi and Zhang, Han and Pang, Jianhua
              and Zhou, Yan",
  journal  = "Sensors (Basel)",
  volume   =  21,
  number   =  4,
  month    =  feb,
  year     =  2021,
  address  = "Switzerland",
  
}

@article{10.1063/1.3491215,
    author = {Dang, H. B. and Maloof, A. C. and Romalis, M. V.},
    title = {Ultrahigh sensitivity magnetic field and magnetization measurements with an atomic magnetometer},
    journal = {Applied Physics Letters},
    volume = {97},
    number = {15},
    pages = {151110},
    year = {2010},
    month = {10},
    issn = {0003-6951},
    doi = {10.1063/1.3491215},
    url = {https://doi.org/10.1063/1.3491215},
    
}

@ARTICLE{10.3389/fphy.2022.946515,
  
AUTHOR={Sato, Katsuaki  and Ishibashi, Takayuki },
         
TITLE={Fundamentals of Magneto-Optical Spectroscopy},
        
JOURNAL={Frontiers in Physics},
        
VOLUME={Volume 10 - 2022},

YEAR={2022},

URL={https://www.frontiersin.org/journals/physics/articles/10.3389/fphy.2022.946515},

DOI={},

ISSN={2296-424X}}

@article{CHEN2023100152,
title = {A review: Magneto-optical sensor based on magnetostrictive materials and magneto-optical material},
journal = {Sensors and Actuators Reports},
volume = {5},
pages = {100152},
year = {2023},
issn = {2666-0539},
doi = {https://doi.org/10.1016/j.snr.2023.100152},
url = {https://www.sciencedirect.com/science/article/pii/S2666053923000152},
author = {Guangyuan Chen and Zhenhu Jin and Jiamin Chen}
}

@article{PhysRevA.82.033807,
  title = {Vector magnetometry based on electromagnetically induced transparency in linearly polarized light},
  author = {Yudin, V. I. and Taichenachev, A. V. and Dudin, Y. O. and Velichansky, V. L. and Zibrov, A. S. and Zibrov, S. A.},
  journal = {Phys. Rev. A},
  volume = {82},
  issue = {3},
  pages = {033807},
  numpages = {7},
  year = {2010},
  month = {Sep},
  publisher = {American Physical Society},
  doi = {10.1103/PhysRevA.82.033807},
  url = {https://link.aps.org/doi/10.1103/PhysRevA.82.033807}
}

@article{PhysRevLett.97.230801,
  title = {High Frequency Atomic Magnetometer by Use of Electromagnetically Induced Transparency},
  author = {Katsoprinakis, G. and Petrosyan, D. and Kominis, I. K.},
  journal = {Phys. Rev. Lett.},
  volume = {97},
  issue = {23},
  pages = {230801},
  numpages = {4},
  year = {2006},
  month = {Dec},
  publisher = {American Physical Society},
  doi = {10.1103/PhysRevLett.97.230801},
  url = {https://link.aps.org/doi/10.1103/PhysRevLett.97.230801}
}

@ARTICLE{Ghaderi_Goran_Abad2021-jc,
  title    = "{Laguerre-Gaussian} modes generated vector beam via nonlinear
              magneto-optical rotation",
  author   = "Ghaderi Goran Abad, Mohsen and Mahmoudi, Mohammad",
  journal  = "Sci Rep",
  volume   =  11,
  number   =  1,
  pages    = "5972",
  month    =  mar,
  year     =  2021,
  address  = "England",
  
}

@article{PhysRevLett.113.013001,
  title = {All-Optical Vector Atomic Magnetometer},
  author = {Patton, B. and Zhivun, E. and Hovde, D. C. and Budker, D.},
  journal = {Phys. Rev. Lett.},
  volume = {113},
  issue = {1},
  pages = {013001},
  numpages = {5},
  year = {2014},
  month = {Jul},
  publisher = {American Physical Society},
  doi = {10.1103/PhysRevLett.113.013001},
  url = {https://link.aps.org/doi/10.1103/PhysRevLett.113.013001}
}

@article{Daloi:22,
author = {Nilamoni Daloi and Tarak Nath Dey},
journal = {Opt. Express},
number = {12},
pages = {21894--21905},
publisher = {Optica Publishing Group},
title = {Vector beam polarization rotation control using resonant magneto optics},
volume = {30},
month = {Jun},
year = {2022},
url = {https://opg.optica.org/oe/abstract.cfm?URI=oe-30-12-21894},
doi = {10.1364/OE.458390},
}

@article{PhysRevA.105.063714,
  title = {Guiding and polarization shaping of vector beams in anisotropic media},
  author = {Daloi, Nilamoni and Kumar, Pardeep and Dey, Tarak Nath},
  journal = {Phys. Rev. A},
  volume = {105},
  issue = {6},
  pages = {063714},
  numpages = {9},
  year = {2022},
  month = {Jun},
  publisher = {American Physical Society},
  doi = {10.1103/PhysRevA.105.063714},
  url = {https://link.aps.org/doi/10.1103/PhysRevA.105.063714}
}

@article{Fabricant_2023,
doi = {10.1088/1367-2630/acb840},
url = {https://doi.org/10.1088/1367-2630/acb840},
year = {2023},
month = {feb},
publisher = {IOP Publishing},
volume = {25},
number = {2},
pages = {025001},
author = {Fabricant, Anne and Novikova, Irina and Bison, Georg},
title = {How to build a magnetometer with thermal atomic vapor: a tutorial},
journal = {New Journal of Physics}
}

@article{Sun:23,
author = {Yujie Sun and Zhaoying Wang},
journal = {Opt. Express},
number = {10},
pages = {15409--15422},
publisher = {Optica Publishing Group},
title = {Optically polarized selective transmission of a fractional vector vortex beam by the polarized atoms with external magnetic fields},
volume = {31},
month = {May},
year = {2023},
url = {https://opg.optica.org/oe/abstract.cfm?URI=oe-31-10-15409},
doi = {10.1364/OE.487426},
}

@article{https://doi.org/10.1002/lpor.202400465,
author = {Cai, Guoan and Tian, Ke and Wang, Zhaoying},
title = {Thermal Atomic Compass Based on Radially Polarized Beam},
journal = {Laser \& Photonics Reviews},
volume = {18},
number = {11},
pages = {2400465},
doi = {https://doi.org/10.1002/lpor.202400465},
url = {https://onlinelibrary.wiley.com/doi/abs/10.1002/lpor.202400465},
year = {2024}
}

@article{Chatterjee:25,
author = {Kalipada Chatterjee and Rajan Jha},
journal = {Opt. Lett.},
number = {10},
pages = {3445--3448},
publisher = {Optica Publishing Group},
title = {Active weak field dynamic magnetometry using conjugated vortex beam interferometry},
volume = {50},
month = {May},
year = {2025},
url = {https://opg.optica.org/ol/abstract.cfm?URI=ol-50-10-3445},
doi = {10.1364/OL.558840},
}

@article{10.1063/1.4923446,
    author = {Shi, Shuai and Ding, Dong-Sheng and Zhou, Zhi-Yuan and Li, Yan and Zhang, Wei and Shi, Bao-Sen},
    title = {Magnetic-field-induced rotation of light with orbital angular momentum},
    journal = {Applied Physics Letters},
    volume = {106},
    number = {26},
    pages = {261110},
    year = {2015},
    month = {06},
    issn = {0003-6951},
    doi = {10.1063/1.4923446},
    url = {https://doi.org/10.1063/1.4923446},

}

@article{Qiu:21,
author = {Shuwei Qiu and Jinwen Wang and Francesco Castellucci and Mingtao Cao and Shougang Zhang and Thomas W. Clark and Sonja Franke-Arnold and Hong Gao and Fuli Li},
journal = {Photon. Res.},
number = {12},
pages = {2325--2331},
publisher = {Optica Publishing Group},
title = {Visualization of magnetic fields with cylindrical vector beams in a warm atomic vapor},
volume = {9},
month = {Dec},
year = {2021},
url = {https://opg.optica.org/prj/abstract.cfm?URI=prj-9-12-2325},
doi = {10.1364/PRJ.418522},
}

@article{Fang:25,
author = {Feiyun Fang and Zhaoying Wang and Qiang Lin},
journal = {Opt. Lett.},
number = {21},
pages = {6457--6460},
publisher = {Optica Publishing Group},
title = {Structured light improves the sensitivity of SERF magnetometry via weak measurement},
volume = {50},
month = {Nov},
year = {2025},
url = {https://opg.optica.org/ol/abstract.cfm?URI=ol-50-21-6457},
doi = {10.1364/OL.573716},
}

@Article{s21041500,
AUTHOR = {Wei, Songrui and Liao, Xiaoqi and Zhang, Han and Pang, Jianhua and Zhou, Yan},
TITLE = {Recent Progress of Fluxgate Magnetic Sensors: Basic Research and Application},
JOURNAL = {Sensors},
VOLUME = {21},
YEAR = {2021},
NUMBER = {4},
ARTICLE-NUMBER = {1500},
URL = {https://www.mdpi.com/1424-8220/21/4/1500},
PubMedID = {33671507},
ISSN = {1424-8220},
DOI = {10.3390/s21041500}
}

@article{10.1063/1.1717911,
    author = {Viehmann, Walter},
    title = {Magnetometer Based on the Hall Effect},
    journal = {Review of Scientific Instruments},
    volume = {33},
    number = {5},
    pages = {537-539},
    year = {1962},
    month = {05},
    issn = {0034-6748},
    doi = {10.1063/1.1717911},
    url = {https://doi.org/10.1063/1.1717911},
    
}

@ARTICLE{133898,
  author={Weinstock, H.},
  journal={IEEE Transactions on Magnetics}, 
  title={A review of SQUID magnetometry applied to nondestructive evaluation}, 
  year={1991},
  volume={27},
  number={2},
  pages={3231-3236},
  doi={10.1109/20.133898}}

@ARTICLE{8425967,
  author={Li, Jundi and Quan, Wei and Zhou, Binquan and Wang, Zhuo and Lu, Jixi and Hu, Zhaohui and Liu, Gang and Fang, Jiancheng},
  journal={IEEE Sensors Journal}, 
  title={SERF Atomic Magnetometer–Recent Advances and Applications: A Review}, 
  year={2018},
  volume={18},
  number={20},
  pages={8198-8207},
  doi={10.1109/JSEN.2018.2863707}}

@book{budker2013optical,
  author    = {Budker, Dmitry and Jackson Kimball, Derek F.},
  title     = {Optical Magnetometry},
  year      = {2013},
  publisher = {Cambridge University Press},
  address   = {Cambridge},
  isbn      = {9780521763227},
  url       = {https://www.perlego.com/book/4219644/optical-magnetometry-pdf}
}

@article{Li:08,
author = {Hebin Li and Vladimir A. Sautenkov and Tigran S. Varzhapetyan and Yuri V. Rostovtsev and Marlan O. Scully},
journal = {J. Opt. Soc. Am. B},
number = {10},
pages = {1702--1707},
publisher = {Optica Publishing Group},
title = {Atomic noise spectra in nonlinear magneto-optical rotation in a rubidium vapor},
volume = {25},
month = {Oct},
year = {2008},
url = {https://opg.optica.org/josab/abstract.cfm?URI=josab-25-10-1702},
doi = {10.1364/JOSAB.25.001702},
}

@article{NBobroff_1993,
doi = {10.1088/0957-0233/4/9/001},
url = {https://doi.org/10.1088/0957-0233/4/9/001},
year = {1993},
month = {sep},
publisher = {},
volume = {4},
number = {9},
pages = {907},
author = {N Bobroff},
title = {Recent advances in displacement measuring interferometry},
journal = {Measurement Science and Technology}
}

@article{10.1117/1.OE.54.10.104104,
author = {Valeriy V. Yashchuk and Nikolay A. Artemiev and Ian Lacey and Wayne R. McKinney and Howard A. Padmore},
title = {{Advanced environmental control as a key component in the development of ultrahigh accuracy ex situ metrology for x-ray optics}},
volume = {54},
journal = {Optical Engineering},
number = {10},
publisher = {SPIE},
pages = {104104},
year = {2015},
doi = {10.1117/1.OE.54.10.104104},
URL = {https://doi.org/10.1117/1.OE.54.10.104104}
}

@misc{CCD_SNR,
  author       = {{Nikon MicroscopyU}},
  title        = {CCD Signal-to-Noise Ratio},
  howpublished = {\url{https://www.microscopyu.com/tutorials/ccd-signal-to-noise-ratio}},
  note         = {Accessed July 30, 2026}
}

@article{2175-ztdj,
  title = {Light storage, retrieval, and controllable interference in an atomic tripod system},
  author = {Zhong, Shan and Sudler, A. J. and Blume, D. and Marino, Alberto M.},
  journal = {Phys. Rev. A},
  volume = {113},
  issue = {1},
  pages = {013103},
  numpages = {16},
  year = {2026},
  month = {Jan},
  publisher = {American Physical Society},
  doi = {10.1103/2175-ztdj},
  url = {https://link.aps.org/doi/10.1103/2175-ztdj}
}

@article{RevModPhys.74.1153,
  title = {Resonant nonlinear magneto-optical effects in atoms},
  author = {Budker, D. and Gawlik, W. and Kimball, D. F. and Rochester, S. M. and Yashchuk, V. V. and Weis, A.},
  journal = {Rev. Mod. Phys.},
  volume = {74},
  issue = {4},
  pages = {1153--1201},
  numpages = {0},
  year = {2002},
  month = {Nov},
  publisher = {American Physical Society},
  doi = {10.1103/RevModPhys.74.1153},
  url = {https://link.aps.org/doi/10.1103/RevModPhys.74.1153}
}

@article{PhysRevA.45.8185,
  title = {Orbital angular momentum of light and the transformation of Laguerre-Gaussian laser modes},
  author = {Allen, L. and Beijersbergen, M. W. and Spreeuw, R. J. C. and Woerdman, J. P.},
  journal = {Phys. Rev. A},
  volume = {45},
  issue = {11},
  pages = {8185--8189},
  numpages = {0},
  year = {1992},
  month = {Jun},
  publisher = {American Physical Society},
  doi = {10.1103/PhysRevA.45.8185},
  url = {https://link.aps.org/doi/10.1103/PhysRevA.45.8185}
}

@misc{Paschotta__rayleigh_length,
  author = {Paschotta, R.},
  publisher = {RP Photonics AG},
  title = {Rayleigh Length},
  year = {},
  howpublished = {RP Photonics Encyclopedia},
  note = {Available online at \url{https://www.rp-photonics.com/rayleigh_length.html}},
  url = {https://www.rp-photonics.com/rayleigh_length.html},
  doi = {10.61835/idz},
  urldate = {2026-07-31}
}

@article{Musha:82,
author = {Toshimitsu Musha and Jun-ichi Kamimura and Masataka Nakazawa},
journal = {Appl. Opt.},
number = {4},
pages = {694--698},
publisher = {Optica Publishing Group},
title = {Optical phase fluctuations thermally induced in a single-mode optical fiber},
volume = {21},
month = {Feb},
year = {1982},
url = {https://opg.optica.org/ao/abstract.cfm?URI=ao-21-4-694},
doi = {10.1364/AO.21.000694},
}

@article{Irie_2008,
doi = {10.1088/0957-0233/19/4/045207},
url = {https://doi.org/10.1088/0957-0233/19/4/045207},
year = {2008},
month = {mar},
publisher = {},
volume = {19},
number = {4},
pages = {045207},
author = {Irie, K and McKinnon, A E and Unsworth, K and Woodhead, I M},
title = {A model for measurement of noise in CCD digital-video cameras},
journal = {Measurement Science and Technology}
}

@article{Sperlich_2014,
doi = {10.1088/0957-0233/25/1/015502},
url = {https://doi.org/10.1088/0957-0233/25/1/015502},
year = {2013},
month = {nov},
publisher = {IOP Publishing},
volume = {25},
number = {1},
pages = {015502},
author = {Sperlich, Karsten and Stolz, Heinrich},
title = {Quantum efficiency measurements of (EM)CCD cameras: high spectral resolution and temperature dependence},
journal = {Measurement Science and Technology}
}

@misc{Steck2021,
  author       = {Daniel A. Steck},
  title        = {Rubidium 87 D Line Data},
  year         = {2021},
  note          = {Revision 2.3.6},
  howpublished = {\url{https://steck.us/alkalidata/rubidium87numbers.pdf}}
}

@article{202400465,
author = {Cai, Guoan and Tian, Ke and Wang, Zhaoying},
title = {Thermal Atomic Compass Based on Radially Polarized Beam},
journal = {Laser \& Photonics Reviews},
volume = {18},
number = {11},
pages = {2400465},
doi = {https://doi.org/10.1002/lpor.202400465},
url = {https://onlinelibrary.wiley.com/doi/abs/10.1002/lpor.202400465},
eprint = {},
year = {2024}
}

@article{Cai:26,
author = {Guoan Cai and Xiaona Liu and Qiang Lin and Zhaoying Wang},
journal = {Photon. Res.},
number = {3},
pages = {649--654},
publisher = {Optica Publishing Group},
title = {Atomic vector magnetic field sensing system based on single radially polarized light},
volume = {14},
month = {Mar},
year = {2026},
url = {https://opg.optica.org/prj/abstract.cfm?URI=prj-14-3-649},
doi = {10.1364/PRJ.573794},
}

@article{Chang:14,
author = {Jintao Chang and Nan Zeng and Honghui He and Yonghong He and Hui Ma},
journal = {Opt. Lett.},
number = {9},
pages = {2656--2659},
publisher = {Optica Publishing Group},
title = {Single-shot spatially modulated Stokes polarimeter based on a GRIN lens},
volume = {39},
month = {May},
year = {2014},
url = {https://opg.optica.org/ol/abstract.cfm?URI=ol-39-9-2656},
doi = {10.1364/OL.39.002656},
}

@article{TIWARI2025109427,
title = {Advances in polarization imaging: Techniques and instrumentation},
journal = {Journal of Quantitative Spectroscopy and Radiative Transfer},
volume = {338},
pages = {109427},
year = {2025},
issn = {0022-4073},
doi = {https://doi.org/10.1016/j.jqsrt.2025.109427},
url = {https://www.sciencedirect.com/science/article/pii/S0022407325000895},
author = {Vipin Tiwari}
}

@article{Patterson2015,
    author  = {Patterson, Luke H. C. and Kihlstrom, Kenneth E. and Everest, Michael A.},
    title   = {Balanced polarimeter: A cost-effective approach for measuring the polarization of light},
    journal = {American Journal of Physics},
    volume  = {83},
    number  = {1},
    pages   = {91--94},
    year    = {2015},
    doi     = {10.1119/1.4896747}
}

@article{Zienau_1975,
doi = {10.1088/0031-9112/26/12/039},
url = {https://doi.org/10.1088/0031-9112/26/12/039},
year = {1975},
month = {dec},
publisher = {},
volume = {26},
number = {12},
pages = {545},
author = {Zienau, S},
title = {Optical Resonance and Two Level Atoms},
journal = {Physics Bulletin}
}

@article{b47h-pdv9,
  title = {Unambiguous vector magnetometry with structured light in atomic vapor},
  author = {Ramakrishna, S. and Fritzsche, S.},
  journal = {Phys. Rev. A},
  volume = {113},
  issue = {3},
  pages = {L031102},
  numpages = {6},
  year = {2026},
  month = {Mar},
  publisher = {American Physical Society},
  doi = {10.1103/b47h-pdv9},
  url = {https://link.aps.org/doi/10.1103/b47h-pdv9}
}

@article{71503,
author = {Choi, Jeonghoon and Kim, Jaekyung and Park, Sujeong and Lee, Doyoon and Baek, Juhoon and Rho, Junsuk and Kim, Minkyung},
title = {Full-Stokes Polarimetry Metasurface-Enabled Single-Shot Measurement of the Spin Hall Effect of Light},
journal = {Advanced Optical Materials},
volume = {n/a},
number = {n/a},
pages = {e71503},
doi = {},
url = {https://advanced.onlinelibrary.wiley.com/doi/abs/10.1002/adom.71503},
eprint = {}
}

@Article{Bhardwaj2025,
author={Bhardwaj, Parkhi
and Dasgupta, Shubhrangshu},
title={Generation of structured light and controlled-NOT gate in microwave regime},
journal={Applied Physics B},
year={2025},
month={Mar},
day={27},
volume={131},
number={4},
pages={90},
issn={1432-0649},
doi={10.1007/s00340-025-08451-w},
url={https://doi.org/10.1007/s00340-025-08451-w}
}

@article{PhysRevLett.127.233202,
  title = {Atomic Compass: Detecting 3D Magnetic Field Alignment with Vector Vortex Light},
  author = {Castellucci, Francesco and Clark, Thomas W. and Selyem, Adam and Wang, Jinwen and Franke-Arnold, Sonja},
  journal = {Phys. Rev. Lett.},
  volume = {127},
  issue = {23},
  pages = {233202},
  numpages = {6},
  year = {2021},
  month = {Nov},
  publisher = {American Physical Society},
  doi = {10.1103/PhysRevLett.127.233202},
  url = {https://link.aps.org/doi/10.1103/PhysRevLett.127.233202}
}

@book{Scully_Zubairy_1997, place={Cambridge}, title={Quantum Optics}, publisher={Cambridge University Press}, author={Scully, Marlan O. and Zubairy, M. Suhail}, year={1997}}

@article{RevModPhys.77.633,
  title = {Electromagnetically induced transparency: Optics in coherent media},
  author = {Fleischhauer, Michael and Imamoglu, Atac and Marangos, Jonathan P.},
  journal = {Rev. Mod. Phys.},
  volume = {77},
  issue = {2},
  pages = {633--673},
  numpages = {0},
  year = {2005},
  month = {Jul},
  publisher = {American Physical Society},
  doi = {10.1103/RevModPhys.77.633},
  url = {https://link.aps.org/doi/10.1103/RevModPhys.77.633}
}

\end{document}